\documentstyle[preprint,prd,eqsecnum,aps,epsfig]{revtex}
\tightenlines
\begin{document}
\newcommand{\h}{\hline}
\newcommand{\Dslashbot}{{/\!\!\!\!D}\hspace{-3pt}_{\bot}}
\newcommand{\Dslash}{/\!\!\!\!D}
\newcommand{\Dbot}{{/\!\!\!\!D}\hspace{-3pt}_{\bot}}
\newcommand{\oDslashbot}{\overleftarrow{/\!\!\!\!D}\hspace{-5pt}_{\bot}}
\newcommand{\oDbot}{\overleftarrow{/\!\!\!\!D}\hspace{-5pt}_{\bot}}
\newcommand{\uvslash}{/\!\!\!\!\hspace{1pt}v}
\newcommand{\vslash}{\not\!{v}}
\newcommand{\kslash}{\not\!{k}}
\newcommand{\qslash}{\not\!{q}}
\newcommand{\omiga}{\omega}
\newcommand{\hr}{\hat{\rho}}

\draft
\title{Inclusive Decays of Bottom Hadrons  \\ 
in New Formulation of Heavy Quark Effective Field Theory}
\author{Y.A Yan, Y.L Wu and W.Y Wang} 
\address{Institute of Theoretical Physics, Chinese Academy of Sciences, 
Beijing 100080, China } 
\date{ylwu@itp.ac.cn}
\maketitle

\begin{abstract}
   We apply the new formulation of heavy quark effective field theory (HQEFT) 
to the inclusive decays of bottom hadrons. The long-term ambiguity of using
heavy quark mass or heavy hadron mass for inclusive decays is clarified within 
the framework of the new formulation of HQEFT. The $1/m_b$ order corrections 
are absent and contributions from  $1/m_b^2$  terms are calculated in detail. 
This enables us to reliably extract the important CKM matrix element $|V_{cb}|$ 
from the inclusive semileptonic decay rates. The resulting lifetime ratios 
$\tau(B^0_s)/\tau(B^0)$ and $\tau(\Lambda_b)/\tau(B^0)$ are found to well agree  
with the experimental data. We also calculate in detail the inclusive semileptonic 
branching ratios and the ratios of the $\tau$ and $\beta$ decay rates as well 
as the charm countings in the $B^0$, $B^0_s$ and $\Lambda_b$ systems. 
For $B^0$ decays, all the observables are found to be consistent with the 
experimental data. More precise data for the $B^0$ decays and further
experimental measurements for the $B^0_s$ and $\Lambda_b$ systems will be 
very useful for testing the framework of new formulation of HQEFT at the level of 
higher order corrections. 
\end{abstract}

\section {Introduction}

  In our previous paper\cite{WWY}, we have provided a more detailed study on 
a new formulation of HQEFT\cite{ylwu} and applied it to evaluate the weak 
transition matrix elements between the heavy hadrons containing a single heavy 
quark. Consequently, the new formulation of HQEFT has exhibited interesting 
features, such as: The Luke's theorem comes out automatically without imposing 
the equation of motion $iv\cdot D Q_{v}^{+} = 0$; the form factors at zero 
recoil are found to be related to the meson masses, so that the most important 
relevant form factors at zero recoil can be fitted from the ground state 
meson masses; the number of universal form factors up to the order of 
$1/m^2_Q$ are less than the one in the usual HQET. 

  It is of interest to apply the new formulation of HQEFT\cite{ylwu} to the 
inclusive 
decays of bottom hadrons. The inclusive decays of bottom hadrons have been 
investigated in the recent years by several groups$\cite{Chay}-\cite{ccc}$. 
While it is well known that in the usual HQET there are still some problems which 
are not yet well understood. These problems mainly involve the following issues: 

  Firstly, the world average values for the lifetime ratios of bottom hadrons 
are \cite{exratio}
\begin{mathletters}
\begin{eqnarray}
{\tau (B^-) \over \tau (B^0)}  = 1.07 \pm 0.03,  \\
{\tau (B^0_s) \over \tau (B^0)}=0.94 \pm 0.04,  \\
{\tau (\Lambda_b) \over \tau (B^0)}=0.79 \pm 0.05. 
\end{eqnarray}
\end{mathletters}
While the usual HQET prediction leads to a uniform lifetime for all the
bottom hadrons when the nonspectator effects are neglected. The lifetime 
differences emerge from $1/m^2_Q$ order which is found to be small. One may 
expect that the $1/m^3_Q$ terms become dominant and provide about $20\%$ 
contribution to the lifetime ratio ${\tau (\Lambda_b)/\tau (B^0)}=0.79 \pm 0.05$ 
(or at least $10\%$ corrections to their total decay rates).
If it is the case, the heavy quark expansion seems to fail in the
inclusive $b$ decays except there are some special reasons to explain
why the $1/m^3_Q$ terms become dorminant and the higher order terms 
$O(1/m^4_Q)$ are smaller. The nonspectator effects have recently been 
considered in ref.\cite{H.Y} and found to result in the following 
predictions  
\begin{mathletters}
\begin{eqnarray}
{\tau (B^-) \over \tau (B^0)} &=& 1.11 \pm 0.02  ,  \\ 
\label{bs}
{\tau (B^0_s) \over \tau (B^0)} &\approx & 1,     \\
{\tau (\Lambda_b) \over \tau (B^0)} & \approx & 0.99 -
       ({f_B \over 185~\mbox{MeV}})^2 (0.007+0.020 \tilde{B})r \geq 0.98 ,
\end{eqnarray}  
\end{mathletters}
where $\tilde{B}$ and $r$ characterize the nonfactorization effects. 
Such a confliction has received wide attention. Authors in 
refs.\cite{Neubert,Baek} have also discussed the lifetime ratios and came to a 
similar result. Though the prediction for $\tau (B^-)/ \tau (B^0)$ agrees with 
the current world average, those of $\tau (B^0_s)/ \tau (B^0)$ and 
$\tau (\Lambda_b)/\tau (B^0)$ deviate somewhat from the central values of the 
world average. In the usual HQET, eq.(\ref{bs}) is the final result because 
the only difference between $B^0$ and $B^0_s$ decays lies in the different CKM 
matrix elements related in the nonspectator effects. 

  To understand the above problems, one of the attempts is to assume that the 
local duality may be violated in the nonleptonic inclusive decays. It was 
suggested in ref.\cite{Altarelli} that a large $1/m_Q$ order correction in 
nonleptonic inclusive decays may exist and be simply described by replacing 
the heavy quark mass by the mass of the decaying hadron in the $m^5$ factor, 
i.e.,
\begin{eqnarray}
\Gamma_{NL} \to \Gamma_{NL} (m_{H_b}/m_b)^5.
\end{eqnarray}

  This assumption was further discussed in refs.\cite{H.Y,9612293,H.Y1}. 
This simple ansatz could not only resolve the lifetime ratio problem, but also 
provide the correct decay widths for the $\Lambda_b$ baryon and the $B$ mesons.
But the charm counting may become much larger than the experimental data.

  Secondly, it seems to have difficulties to simultaneously explain the 
semileptonic branching ratio $B_{SL}$ and the charm counting $n_c$ in $B^0$ 
decays. In general,  when the charm counting is required to be near the 
experimental data, the predicted semileptonic branching ratio in the usual HQET 
is significantly larger than the experimental data. It has been shown in 
ref.\cite{Bagan} that for $m_b/2 \leq \mu \leq 2m_b$ and 
$m_{b} = 4.8\pm 0.2$GeV 
\begin{mathletters}
\begin{eqnarray}
   B_{SL} &=& (12.0 \pm 0.7  \pm 0.5 \pm 0.2^{+0.9}_{-1.2})\%  , \\
n_c  &= & 1.24\mp 0.05\pm 0.01, \\
\bar{B}_{SL} &=& (11.3 \pm 0.6  \pm 0.7 \pm 0.2^{+0.9}_{-1.7})\% ,  \\
\bar{n}_c  &= & 1.30\mp 0.03\pm 0.03 \pm 0.01 ,
\end{eqnarray}
\end{mathletters}
where the first result is for the OS scheme and the second one for 
$\overline{MS}$ scheme. After considering the spectator effects, the results 
has been found to be improved \cite{Neubert} 
\begin{mathletters}
\begin{eqnarray}
   B_{SL} &=& \cases{
    12.0\pm 1.0\%  ;& $\mu=m_b$, \cr
    10.9\pm 1.0\%  ;& $\mu=m_b/2$, \cr} \\
   \phantom{ \bigg[ }
   n_c  &=& \cases{
    1.20\mp 0.06   ;& $\mu=m_b$, \cr
    1.21 \mp 0.06  ;& $\mu=m_b/2$. \cr}
\end{eqnarray}
\end{mathletters}
It is seen that the uncertainties in the two quantities are anti-correlated.
One may compare them with the world average\cite{pdg}
\begin{mathletters}
\begin{eqnarray}
&& Br(b\to c e \bar{\nu})=10.48 \pm 0.5\% ,  \\
&& n_c=1.17 \pm 0.04. 
\end{eqnarray}
\end{mathletters}

  Thirdly, there is an inconsistent picture between the Luke's theorem\cite{Luke}   
for the exclusive heavy hadron decays and the Chay-Georgi-Grinstein 
theorem\cite{Chay} for the inclusive heavy hadron decays. Luke's theorem tells 
us that the $1/m_Q$ order corrections are absent if one uses the meson mass
to normalize the weak matrix elements in the exclusive heavy to heavy transitions 
at zero recoil. While according to the Chay-Georgi-Grinstein theorem, $1/m_Q$ order 
corrections are absent only  when the quark mass is used. Consequently, 
the prediction of total decay width strongly depends on the value of quark mass.

  These problems may arise from the simple treatment in the usual HQET, where 
the bound state effects and hadronization have not been taken into account. 
This may be a strong indication for a necessity of developing a new formulation 
of effective theory to incorporate these effects. For this purpose, we are going 
to devote in this paper the investigation of the inclusive bottom hadron decays 
and to provide better understanding on the problems mentioned above within the 
framework of the new formulation of HQEFT\cite{ylwu}. 
In section II, we briefly describe the framework of the new formulation of 
HQEFT. In section III, we apply it to the inclusive bottom hadron decays 
$ H_{b} \to X_{c} \ell \nu$  and present a general formulation 
for $ b \to c$ transitions via the heavy quark expansion. In section IV, we 
investigate the inclusive decays of $B^0$ and $B^0_s$ mesons as well as 
$\Lambda_b$ baryon by providing detailed numerical results for their 
semileptonic branching ratios and lifetime ratios as well as the charm 
countings $n_c$. It is interesting to see that the results obtained 
by using the new formulation of HQEFT are consistent with the experimental 
data. In particular, the new formulation of HQEFT allows us to simply clarify 
the well known ambiguity of using the quark mass or hadron mass in the 
inclusive heavy hadron decays. As a consequence, 
the CKM matrix element $|V_{cb}|$ is well determined from the semileptonic 
decay rate. Conclusions and remarks are presented in the last section.

\section {New Formulation of HQEFT}

  For completeness, we present in this section a brief description on the 
framework of the new formulation of HQEFT\cite{WWY,ylwu}. 
Let us begin with the effective Lagrangian for a heavy quark 
\begin{eqnarray}
\label{Leff}
{\cal L}_{eff} &=& \bar{Q}^{(+)}_v iv \cdot D Q^{(+)}_v+{1 \over m_Q}
     \bar{Q}^{(+)}_v(i \Dbot)^2 Q^{(+)}_v  
\nonumber \\
&&   -{1 \over 2m^2_Q} \bar{Q}^{(+)}_v i \Dbot iv \cdot D i \Dbot Q^{(+)}_v 
 \\
&&      +{1 \over (2m_Q)^2} \bar{Q}^{(+)}_v (i \Dbot)^2 {1 \over iv \cdot D}
     (i \Dbot)^2 Q^{(+)}_v+O(1/{m^3_Q})  \\
&=&{\cal L}^{(0)}_{eff} + {\cal L}^{(1/m_Q)}_{eff} ,
\nonumber
\end{eqnarray}
where $\overleftarrow{D}_\mu$ and $\Dbot^{\ \mu}$ are defined as
$$\chi \overleftarrow{D}_\mu \equiv \partial_\mu \chi+i\chi g A_\mu^a T^a;
\quad \Dbot^{\ \mu} \equiv (g^{\mu\nu} -v^\mu v^\nu) D_\nu$$
and  ${\cal L}^{(0)}_{eff}=\bar{Q}^{(+)}_v iv \cdot D Q^{(+)}_v$ 
denotes the leading term and 
${\cal L}^{(1/m_Q)}_{eff}$  represents the terms suppressed by the powers 
of $1/m_Q$   
.
And the representation of left-handed current 
$J^\mu=\bar{Q}^\prime \Gamma^\mu Q$ is
\begin{eqnarray}
\label{Jeff}
J^\mu_{eff} &=& e^{i(m_{Q^\prime} v^\prime - m_Q v) \cdot x}
   \bar{Q}^{\prime {(+)}}_{v^\prime} \{\Gamma^\mu  
   +{1 \over {2m_{Q^\prime}}}(i\oDbot\hspace{0.5pt}^\prime)^2 
   {1 \over {-i v^\prime \cdot \overleftarrow{D}}} \Gamma^\mu
\nonumber \\
&&+{1 \over {2m_Q}}\Gamma^\mu {1 \over {i v \cdot D}}(i\Dbot)^2 \} Q^{(+)}_v 
   +O(1/m^2_Q) ,
\end{eqnarray}
with $\Gamma^\mu=\gamma^\mu {1-\gamma^5 \over 2}$
and $D^{\prime\mu}_\bot=(g^{\mu\nu}-v^{\prime\mu} v^{\prime\nu}) D_\nu$.

   In order to exhibit a manifest spin-flavor symmetry in the HQEFT, it is 
useful to introduce a hadron state $|H_v>$ corresponding to the effective heavy 
quark field $Q^{(+)}_v$. The hadron state $|H_v>$ is related to the state $|H>$ 
of heavy hadron $H$ by the following equation
\begin{equation}
\label{hadronMatrix}
<H^{\prime}|J^\mu |H>= \sqrt{\frac{m_{H^\prime} m_H}{\bar{\Lambda}_{H^\prime} 
	\bar{\Lambda}_H}} <{H^\prime}_{v^\prime}| {J^\mu}_{eff} 
	e^{i\int d^4x ({\cal L}^{(1/m_Q)}_{eff}+
	{\cal L}^{(1/m_{Q^\prime})}_{eff})}|H_v> ,
\end{equation}  
  with $m_H$ being the mass of the heavy hadron $H$ and $ \bar{\Lambda}_H
=m_H-m_Q$. The factor $m_H$ comes from the standard normalization
\begin{eqnarray}
\label{normInQCD}
&&<H(p^\prime)|H(p)>= 2 p^0 (2\pi)^3 \delta^3({\bf p-p^\prime})  .
\end{eqnarray}
Where the factor $\sqrt{\bar{\Lambda}_{H^\prime} \bar{\Lambda}_H}$ is 
introduced to make the normalization concerning the hadron state $|H_v>$ to be 
independent of the heavy flavor, i.e.,
\begin{eqnarray}
<H_v | \bar{Q}^{(+)}_v \gamma^\mu Q^{(+)}_v |H_v> &=& 2\bar{\Lambda} v^\mu ,
\end{eqnarray}    
   with
$$\bar{\Lambda}= \lim_{m_Q \rightarrow \infty} \bar{\Lambda}_H .$$

  Expanding the right-hand side of eq.(\ref{hadronMatrix}), to the order of 
$1/m_Q$, we have 
\begin{eqnarray}
<H^\prime|J^\mu|H> &=& \sqrt{m_H m_{H^\prime} \over \bar{\Lambda}_H 
	\bar{\Lambda}_{H^\prime}} 
       <H^\prime_{v^\prime}| \bar{Q}^{\prime (+)}_{v^\prime} \{ \Gamma^\mu
-{1 \over 2m_{Q^\prime}} (i\overleftarrow{D^\prime}{\hspace{-5pt}_{\bot}})^2 
	{1 \over -iv^\prime \cdot \overleftarrow{D}} \Gamma^\mu
\nonumber  \\
&& -{1 \over 2m_Q} \Gamma^\mu {1 \over iv \cdot D} (i \Dslashbot)^2
	+O({1 \over m^2_{Q^{(\prime)}}}) \}Q^{(+)}|H_v> .
\end{eqnarray}

 In general, a heavy quark in a hadron cannot 
truly be on-shell due to strong interactions among heavy quark and  
light quark as well as soft gluons. The off-shellness of heavy quark in the 
hadron is characterized by a residual momentum $k$. The total momentum $P_Q$ 
of the heavy quark in a hadron may be written as:  $P_Q=m_Q v + k$. In the 
usual HQET one mainly deals with the heavy quark and treats the light quark as 
a spectator, which does not affect the properties of heavy hadrons to a large 
extent (except the effects of weak annihilation and Pauli interference). 
However, since the light degrees of freedom affect the character of heavy 
hadrons, it may be this simple treatment that meets difficulties in explaining 
the lifetime differences between $B^0$ and $B^0_s$ as well as between 
$\Lambda_b$ and $B$ as mentioned in the introduction. To take the effects of 
light degrees into account and not to deal with hadronization directly, 
we will adopt an alternative picture, namely the residual 
momentum $k$ of the heavy quark within a hadron is considered to comprise the 
contributions of the light degrees of freedom. With this picture the heavy quark 
may be regarded as a `dressed heavy quark', thus the 
heavy hadron containing a single heavy quark may be more reliable to be 
considered as a dualized particle of a `dressed heavy quark'. 
Thus the momentum $P_H$ of a hadron $H$ is decomposed into 
\begin{eqnarray}
P_H=m_Q v + k + k^\prime ,
\end{eqnarray}
with $k^\prime$ being the momentum depending on the heavy flavor and suppressed
by $1/m_Q$. With this picture, the momentum of the `dressed heavy quark'  
inside the hadron is given by 
\begin{eqnarray}
P_Q = \lim_{m_Q \rightarrow \infty} P_H.
\end{eqnarray}
  Hence
\begin{equation}
P^2_H=m^2_H =m^2_Q +2m_Q v \cdot (k+k^\prime)+ k^{\prime 2} +k^2 +2k \cdot k^\prime .
\end{equation}

  Then
\begin{eqnarray}
\bar{\Lambda}_H & \equiv & m_H - m_Q
=  {v \cdot (k+k^\prime) \over 1+ \bar{\Lambda}/2m_Q} + {(k+k^\prime)^2 \over 2m_Q 
	(1+ \bar{\Lambda}/2m_Q)} ,
\end{eqnarray}
and
\begin{equation}
\label{deflambda}
\bar{\Lambda} = \lim_{m_Q \rightarrow \infty}\bar{\Lambda}_H = 
\lim_{m_Q \rightarrow \infty} v \cdot (k+k^\prime)
= v \cdot k .
\end{equation}

  In terms of the operator formulation, it implies that
\begin{equation}
\label{motion}
<H_v| \bar{Q}^{(+)}_v i v \cdot D Q^{(+)}_v |H_v>=\bar{\Lambda} 
	<H_v| \bar{Q}^{(+)}_v Q^{(+)}_v |H_v> .
\end{equation}
Thus to simplify the evaluation of the matrix elements, one 
may approximately replace the propagator $1/iv \cdot D$ by $1/\bar{\Lambda}$ 
\begin{eqnarray}
\label{lambda}
\kappa_1 &\equiv & -<H_v|\bar{Q}^{(+)}_v D_\bot^2 
	Q^{(+)}_v|H_v>/(2\bar{\Lambda})  , \nonumber \\
\kappa_2 &\equiv & <H_v|\bar{Q}^{(+)}_v g \sigma_{\mu\nu} 
G^{\mu\nu} Q^{(+)}_v|H_v>/(4d_H \bar{\Lambda})  .
\end{eqnarray}
where $d_H = -3$ for pseudo scalar mesons, $d_H = 1$ for vector mesons and
$d_H = 0$ for ground state heavy baryons.

  With this approximation, the mass formulae can be simply 
given in terms of the effective Lagrangian ${\cal L}^{(1/m_Q)}_{eff}$
\begin{eqnarray}
\label{massOfL}
m_H &=& m_Q + \bar{\Lambda} - {<H_v| \bar{Q}^{(+)}_v (i \Dslashbot)^2 
      Q^{(+)}_v |H_v> \over 2\bar{\Lambda} \cdot m_Q}  +O({1 \over m^2_Q})
\nonumber  \\
& \approx & m_Q + \bar{\Lambda} - <H_v| {\cal L}^{(1/m_Q)}_{eff} |H_v>/
	(2\bar{\Lambda})   ,
\end{eqnarray}
 which can be reexpressed as follows by using eq.(\ref{lambda})
\begin{eqnarray}
\label{massl1l2}
m_H \approx m_Q + \bar{\Lambda} -{\kappa_1 \over m_Q} + 
	{d_H \kappa_2 \over m_Q} + O({1 \over m^2_Q}) .
\end{eqnarray}
It is useful to define the mass of `dressed heavy quark' as 
\begin{equation}
\label{addmass0}
\hat{m}_Q \equiv \lim_{m_{Q}\to \infty} m_{H} = m_Q +\bar{\Lambda},
\end{equation}
which can be expressed in terms of the hadron mass 
\begin{eqnarray}
\label{addmass}
\hat{m}_Q & = & m_H+{\kappa_1-d_H \kappa_2 \over m_Q} +O({1 \over m^2_Q})   
    \nonumber  \\
&=& m_H+{\kappa_1-d_H \kappa_2 \over m_H} +O({1 \over m^2_H}).
\end{eqnarray}

\section {Dynamics of Inclusive Decays}

  The techniques for inclusive decays of heavy hadrons in the framework of
effective theory was developed in the early years of 
this decade. Here we shall extend the method of ref.\cite{Wise} to the 
framework of new formulation of HQEFT. Let us first briefly recall the basic 
formulae for the description of inclusive semileptonic decay
$$H(P_H=m_H v) \to X_c(P_X) +\ell (p)+\bar{\nu}_\ell (p^\prime); 
       \quad q=p+p^\prime ,$$
which is mediated by the effective Hamiltonian
\begin{eqnarray}
  {\cal H}_{eff}={4G_F \over \sqrt{2}} V_{cb} \bar{b} \Gamma^\mu c 
   \bar{\nu}_\ell \Gamma_\mu \ell
 ={4G_F \over \sqrt{2}} V_{cb} J^\mu J_{\ell \mu} .
\end{eqnarray}

   To calculate the decay rates, it is useful to introduce the hadronic tensor
\begin{eqnarray}
\label{hadten}
W^{\mu\nu} ={1 \over 2m_H} (2\pi)^3{\sum\limits_{X_c}} \delta^4(P_{H_b}-q-P_X)
      <H|J^{\mu\dag}|X_c><X_c|J^\nu|H> ,
\end{eqnarray}
which can be expanded in terms of five form factors when one averages
the spin of initial and final states
\begin{eqnarray}
W^{\mu\nu} = -g^{\mu\nu} W_1 + v^\mu v^\nu W_2 -i \epsilon^{\mu\nu\alpha\beta}
 v_\alpha q_\beta W_3
+q^\mu q^\nu W_4+(q^\mu v^\nu + q^\nu v^\mu) W_5 .
\end{eqnarray}

  For massless lepton pair, $W_4$ and 
$W_5$ will not contribute to the decay width 
\begin{eqnarray}
\label{qxy}
 {d\Gamma \over dq^2 dE_e dE_\nu}={G^2_F V^2_{cb} \over 2 \pi^3} \Theta(E_e-q^2
  /4E_\nu)\{q^2 W_1 + {1 \over 2}(E_e E_\nu-q^2) W_2 + q^2 (E_e-E_\nu) W_3 \}. 
\end{eqnarray}

  To evaluate the form factors $W_i$, let us consider the time ordered product
\begin{eqnarray}
\label{Tmunu}
T^{\mu\nu}&=& -{i \over 2m_H} \int d^4x e^{-iq \cdot x} <H| {\cal T} 
   \{ J^{\mu\dag} (x) J^\nu (0) \} |H>   \\
&=& -g^{\mu\nu} T_1 +v^\mu v^\nu T_2 -i \epsilon^{\mu\nu\alpha\beta} v^\alpha 
   q^\beta T_3+q^\mu q^\nu T_4 +(v^\mu q^\nu+v^\nu q^\mu) T_5 .  \nonumber
\end{eqnarray}
It is known that the form factors $W_i$ are related to $T_i$ via
\begin{eqnarray}
\label{TiWi}
W_i=-{1 \over \pi} Im T_i  .
\end{eqnarray}

  Explicitly, the quark matrix element in eq.(\ref{Tmunu}) is given by
\begin{eqnarray}
\label{qm}
{1 \over 2m_H} <H|\bar{b} {\gamma^\mu (m_b \vslash+\kslash-\qslash) \gamma^\nu
  \over (m_b v +k -q)^2 -m^2_c+i\epsilon}{1-\gamma_5 \over 2} b |H> .
\end{eqnarray}
Here $m_b v +k -q = P_c$ is the momentum of charm quark.

  Within the framework of new formulation of HQEFT, as we have discussed above, 
the residual momentum of the bottom quark in the hadron is given by 
$v \cdot k \sim \bar{\Lambda}$. For ensuring the leading term to have the 
largest contribution, we may expand eq.(\ref{qm}) in terms of the small 
subtracted momentum  $$k-v <v \cdot k>.$$
Here
\begin{eqnarray}
<v \cdot k>  \equiv <H|\bar{b} e^{-im_b \uvslash v \cdot x} iv \cdot D 
e^{i m_b \uvslash v \cdot x} b |H>/2m_H	 = {\bar{\Lambda} }.
    \nonumber
\end{eqnarray}

  To evaluate the matrix element eq.(\ref{qm}), we need to know the following
relevant matrix elements
\begin{mathletters}
\begin{eqnarray}
&&{1 \over 2m_H} <H|\bar{b} b |H> =1  ,  \\
&&{1 \over 2m_H} <H|\bar{b} e^{-im_b \uvslash v \cdot x} 
{1 \over 2} (i D^{\kappa}_\bot iD^{\tau}_\bot + i D^{\tau}_\bot iD^{\kappa}_\bot)
  e^{i m_b \uvslash v \cdot x} b |H> 
  =A (g^{\kappa\tau} - v^\kappa v^\tau)  ,\\
&&{1 \over 12m_H} <H| \bar{b} g \sigma_{\mu\nu} G^{\mu\nu} b |H> 
     = N_b .
\end{eqnarray}
\end{mathletters}
Up to the leading order in $1/m_b$, we have, from
eqs.(\ref{motion}) and (\ref{lambda}), the following results for $A$ and $N_{b}$
\begin{mathletters}
\begin{eqnarray}
A &=& {\kappa_1 \over 3 } ,\\
N_b &=& {d_H \kappa_2 \over 3 }   .
\end{eqnarray}
\end{mathletters}

  The matrix element concerning one gluon in eq.(\ref{Tmunu}) has the form 
\begin{eqnarray}
\label{gm}
<H|\bar{b} {g \over 2} G^{\alpha\beta} 
     \epsilon_{\alpha\beta\kappa\sigma} {(m_b v +k-q)^\kappa \over 2 m_H}
{g^{\mu\sigma} \gamma^\nu+ g^{\nu\sigma} \gamma^\mu -g^{\mu\nu} \gamma^\sigma 
    +i \epsilon^{\mu\sigma\nu\tau} 
\over [(m_b v +k -q)^2 -m^2_c+ i \epsilon]^2} \Gamma^\tau b |H> .  
\end{eqnarray}

  Expanding eqs.(\ref{qm}) and (\ref{gm}), one can extract the form factors $T_i$.
Using the relation eq.(\ref{TiWi}), one then obtains $W_i$.
Here we shall only list the final results for $W_i$
\begin{mathletters}
\label{Wi}
\begin{eqnarray}
W_1 &=& {1 \over 2}(\hat{m}_b-E_e-E_\nu) \delta (z) 
     + \{{1 \over 2}(3A+N_b)(\hat{m}_b-E_e-E_\nu) \} \delta^\prime (z)
\nonumber  \\
&& +A (\hat{m}_b-E_e-E_\nu) \{ q^2-(E_e+E_\nu)^2 \} \delta^{\prime\prime}(z) ,
   \\
W_2 &=& \hat{m}_b \delta (z) + \{\hat{m}_b (3A-N_b)+2 A (E_e+E_\nu) \} 
  \delta^\prime (z)
\nonumber  \\
&& + 2 \hat{m}_b A \{ q^2-(E_e+E_\nu)^2 \} \delta^{\prime\prime}(z)  ,\\
W_3 &=& {1 \over 2} \delta (z) - \{ {1 \over 2} (5 A+N_b) \} \delta^\prime (z)
+A \{ q^2-(E_e+E_\nu)^2 \} \delta^{\prime\prime}(z)  ,
\end{eqnarray}
\end{mathletters}
with $z=(\hat{m}_b v -q)^2-m_c^2=
 \hat{m}_b^2+q^2-2\hat{m}_b E_e-2\hat{m}_b E_\nu-m_c^2$. 

  Substituting eq.(\ref{Wi}) into eq.(\ref{qxy}) and integrating out the 
variables $E_\nu$ and $q^2$, one yields
the inclusive lepton spectrum
\begin{eqnarray}
{1 \over \hat{\Gamma}_0} {d\Gamma \over dy} &=&-2\rho^2(3-\rho)
   +{4\rho^3 \over (1-y)^3}
   -{6\rho^2 (1+\rho) \over (1-y)^2} +{12\rho^2 \over 1-y} +6(1-\rho)y^2-4y^3
\nonumber  \\
&&+{A \over \hat{m}^2_b} \{-6\rho^3+ 12\rho^2 -{24\rho^3 \over (1-y)^5}
     +{6\rho^2(3+5\rho) \over (1-y)^4} 
  	-{12\rho^2 \over (1-y)^3} -{18\rho^2 \over (1-y)^2} +6y^2 \}
\nonumber   \\
&&+{N_b \over \hat{m}^2_b} \{ -6\rho (2+\rho) -{12\rho^2 \over (1-y)^3}+{6\rho 
   (2+3\rho) \over (1-y)^2}-24 \rho y -18 y^2 \},
\end{eqnarray}
with  
\begin{eqnarray}
 \hat{\Gamma}_0 & \equiv & {G_F^2 \hat{m}_b^5 V_{cb}^2 \over 192 \pi^3} = 
{G_F^2 m_H^5 V_{cb}^2 \over 192 \pi^3} \left( 1 + {\kappa_1-d_H \kappa_2 \over m_H^2}
 +O({1 \over m^3_H}) \right)^{5} \nonumber \\
& \equiv & \Gamma_{0} \left( 1 + {\kappa_1-d_H \kappa_2 \over m_H^2}
 +O({1 \over m^3_H}) \right)^{5}.
\end{eqnarray}  

Where $y$ and $\rho$ are rescaled variables defined as
\[
y \equiv 2 E_e/\hat{m}_b, \quad \rho \equiv m_c^2/\hat{m}_b^2. \]

The kinetic region for $y$ is
\begin{eqnarray}
 0 \leq y \leq {P_b^2-P_c^2 \over P_b^2} - {E_\nu \over 2 v \cdot P_b}.
\end{eqnarray}
Treating charm quark $c$ as heavy quark and using $v \cdot k=\bar{\Lambda}$,
one has
\begin{eqnarray}
 0 \leq y \leq  1-\hat{\rho},		
\quad \hat{\rho}\equiv \hat{m}^2_c/\hat{m}^2_b.
\end{eqnarray}
  Thus the total decay width for $b \to c e \bar{\nu}$ is found to be
\begin{eqnarray}
\label{Gsl}
{\Gamma \over \hat{\Gamma}_0} &=& -(1-\hat{\rho})^4+2(1-\hat{\rho})^3 (1-\rho)
   -2 \rho^2 (1-\hat{\rho})(3-\rho)  \nonumber  \\
&&   -2\rho^3 (1-{1\over \hat{\rho}^2} +6 \rho^2 (1+\rho) 
    (1-{1 \over \hat{\rho}}) -12 \rho^2 \ln\hat{\rho}  
\nonumber  \\
&&   +2{A\over \hat{m}^2_b \hat{\rho}^4} (1-\hat{\rho})^3 
   \{\hr^4 + \hr \rho^2 (3-4\rho) + 3\hr^2 \rho^2 (2 - \rho) -3 \rho^3\}
\nonumber  \\
&&   -{6 N_b \over \hat{m}^2_b \hr^2} (1-\hr)^3 (\hr-\rho)^2 .
\end{eqnarray}

  Similarly, we find that $b\to c$ transitions have the following general forms
\begin{mathletters}
\begin{eqnarray}
\label{sl}
{1 \over \hat{\Gamma}_0} \Gamma (b \to c\ell\nu)&=& \eta_{c\ell}(\rho,\rho_\ell,\mu) 
    \{ I_0(\rho,\rho_\ell,\hr)+I_1(\rho,\rho_\ell,\hr) {A \over \hat{m}^2_b} 
    	+I_2(\rho,\rho_\tau,\hr){N_b\over \hat{m}^2_b} \}   ,  \\
{1\over \hat{\Gamma}_0} \Gamma (b \to cud^\prime)  &=& 3 \eta_{cu}(\rho,\mu) 
      \{ I_0 (\rho,0,\hr) + 
	I_1 (\rho,0,\hr) {A \over \hat{m}^2_b} 
\nonumber  \\ 
  &&+I_2(\rho,0,\hr){N_b\over \hat{m}^2_b} \}
  - ( c^2_{+}(\mu) -c^2_{-}(\mu)) 6 (1-\rho)^3 \kappa_2/\hat{m}^2_b  ,\\
{1\over \hat{\Gamma}_0} \Gamma(b \to ccs^\prime)  &=& 3 \eta_{cc}(\rho,\mu) 
  \{ I_0(\rho,\rho,\hr) +
	I_1(\rho,\rho,\hr) {A \over \hat{m}^2_b} 
\nonumber  \\
   &&+I_2(\rho,\rho,\hr){N_b\over \hat{m}^2_b} \}
   - (c^2_{+}(\mu)-c^2_{-}(\mu)) I_3(\rho,\rho,0) 6\kappa_2/\hat{m}^2_b ,
\end{eqnarray}
\end{mathletters}
with $s^\prime = d V_{cd}+s V_{cs}$ and $d^\prime =d V_{ud} +s V_{us}$
and $\rho_\ell \equiv m_\ell^2/\hat{m}_b^2$. 
The $\eta$ functions arise from QCD radiative corrections. The one-loop results 
have been calculated in refs.\cite{Bagan,MBV,AC}. Here we will use the two-loop 
results in refs.\cite{MM,LLSS} and adopt the reference value $m_c/m_b = 0.3$, 
which leads to the following results
\begin{mathletters}
\begin{eqnarray}
\eta_{ce} &=& 1 - 0.53 \alpha_s(\mu) - 1.53 \alpha_s(\mu)^2 , \\
\eta_{c\tau} &=& 1 - 0.44 \alpha_s(\mu) - 1.44 \alpha_s(\mu)^2 , \\
\eta_{cu} &=&     1 + 0.95 \alpha_s(\mu) + {\alpha_s(\mu)^2 \over \pi^2} [33.03 + 
    3.34 \ln (m_W/m_b) + 4 \ln (m_W/m_b)^2] , \\
\eta_{cc} &=&  1 - 0.21 \alpha_s(\mu) + {\alpha_s(\mu)^2 \over \pi^2} 
      [-9.99 + 7.17 \ln (m_W/m_b) + 4 \ln (m_W/m_b)^2] .
\end{eqnarray}
\end{mathletters}
The functions $I_i(x,y,0)$ are phase space factors at tree level. 
Their explicit forms are
\begin{mathletters}
\begin{eqnarray}
I_0(x,y,z) &=& \sqrt{\kappa} \{ 1 - z - {{z}^2} + {{z}^3} -
  2\,{{z}^2}\,x + 2\,{x^3} -
  2\,x\,{{ ( -1 + y  ) }^2} \nonumber  \\
&& -{\frac{6\,{{x}^2}\,{{ ( -1 + y  ) }^2}}{z}}
  + {\frac{2\,{x^3}\,
      {{ ( -1 + y  ) }^2}}{{z^2}}} -
  7\,y - {z^2}\,y -
  7\,{y^2} - z\,{y^2}  \nonumber  \\
&&+{y^3} + 4\,z\,x\,
    ( 1 + y  )  -
  6\,{x^2}\, ( 1 + y  )  -
  {\frac{4\,{x^3}\, ( 1 + y  ) }
    z} \}
\nonumber  \\
&& +12 x^2 \ln {(1+z-y+\sqrt{\kappa})^2 \over 4z} +12 y^2 \ln 
  {(1-z+y+\sqrt{\kappa})^2 \over 4y} 
\nonumber  \\
&&  -12 x^2 y^2 \ln {(1-z-y+\sqrt{\kappa})^2 \over 4zy}  ,\\
I_1(x,y,z) &=& {2 \sqrt{\kappa} \over z^4} \{ {z^6} -
       3\,{x^3}\,{{ ( -1 + y ) }^4} -
       2\,{z^5}\, ( 1 + y )  +
       {z^3}\,{x^2}\,( -9 + 2\,x	 \nonumber  \\
&& - 18\,y + 12\,x\,y + 39\,{y^2} )  
 + {z^4}\,[ -3\,{x^3} +
          {{( -1 + y ) }^2} +
          3\,{x^2}\,( 2 + 3\,y ) ]  \nonumber  \\
&& + z\,{x^2}\,{{( -1 + y) }^2}\,
        [ 3\,{{ ( -1 + y  ) }^2} +
          2\,x\,  ( 1 + 6\,y  )  ]  \nonumber  \\
&& +{z^2}\,{x^2}\,
        [ -15\,{{ ( -1 + y  ) }^2}\,
           y + x\,
            ( 2 + 8\,y - 18\,{y^2}  )] \} 
\nonumber  \\
&&  +72 x^2 y^2 \ln {(1-z-y+\sqrt{\kappa})^2 \over 4zy}  ,\\
I_2(x,y,z) &=& -6(x-z)^2 [ z^2 +(1-y)^2 -2 z (1+y) ]^{3\over 2},    \\
I_3(x,x,0) &=& \sqrt{1-4x} (1+{x \over 2}+3x^2)-3x(1-2x^2) \ln 
  {1+\sqrt{1-4x} \over 1-\sqrt{1-4x}},
\end{eqnarray}
\end{mathletters}
with $\kappa= (1+y-z)^2-4y.$ Note that for $z=x$, $I_0$ agrees with that
in refs.\cite{Wise,Falk}.
  
  Before ending this section, we would like to address the following points:
Firstly, the mass entering into the factor $\hat{\Gamma}_0$ is neither the
heavy quark mass $m_b$ nor the hadron mass $m_H$, it is the so-called
`dressed heavy quark' mass $\hat{m}_b=m_b + \bar{\Lambda} = 
m_{H_b} (1+ O(1/m^2_{H_b}))$, which differs from the hadron mass by terms 
suppressed by $1/m_b$. Secondly, paralleling to Luke's theorem, there is no 
$1/m_b$ order correction when the hadron mass is used. Thus the uncertain 
parameters, i.e., the bottom quark mass $m_b$ and binding energy 
$\bar{\Lambda}$, do not enter separately into the expression of decay widths. 
Furthermore, one may notice that for the final charm quark, 
both charm quark mass $m_{c}$ and `dressed charm quark' mass $\hat{m}_{c}$ have 
entered into the phace space factors of the general formulation of the decay rates. 
Where the quark mass $m_{c}$ comes from the propagator of charm quark, 
and the `dressed charm quark' mass $\hat{m}_{c}$ arises from the momentum of 
charm quark inside the hadron, i.e., $P_{c}= P_{b} - q = \hat{m}_{b}v - q = 
\hat{m}_{c}v' $, for a parallel treatment to the bottom quark inside the hadron. 
These features enable us to determine the lifetime ratios, semileptonic branching 
ratios and even the lifetimes of bottom hadrons in rather accuracy.

\section {Numerical Results and Discussions}

 In our numerical calculations, we shall neglect the contributions from penguin 
diagrams and other rare decays and incorporate the two-loop radiative 
corrections in refs.\cite{MM,LLSS} to calculate the values of observable 
parameters in $b$ decays. Such an approximation causes errors no more than 2\%. 
Thus the total decay width for bottom hadrons are given by
\begin{eqnarray}
\Gamma^t_H = {1 \over \tau(H)} \simeq 
  \Gamma (b\to c u d^\prime) + \Gamma (b \to c c s^\prime)+
     \sum\limits_{\ell} \Gamma (b\to c \ell \bar{\nu})+\Gamma(b \to u) .
\end{eqnarray}
The semileptonic branching ratios are defined as    
$$B_{SL}(H) \equiv Br(H_b\to X_c e \bar{\nu})=
 {\Gamma(H_b \to X_c e \bar{\nu}) \over \Gamma^t_H} ,$$
$$B_\tau(H) \equiv Br(H_b\to X_c \tau \bar{\nu})=
 {\Gamma(H_b \to X_c \tau \bar{\nu}) \over \Gamma^t_H} .$$
Other two ratios concerning the charm counting
  $$n_c(H)=1+{\Gamma  (H_b \to X_{c\bar{c}}) \over \Gamma^t_H }-
             {\Gamma_{~\mbox{nocharm}} \over \Gamma^t_H}$$
and relative leptonic contributions between the $\tau$ 
decay $B_\tau(H)$ and the $\beta$ decay $B_{SL}(H)$ 

$$ R(H) \equiv {B_\tau(H) \over B_{SL}(H)}.$$

  Note that the formula given above are general and can be applied to bottom 
hadrons by just taking different binding energies $\bar{\Lambda}(H)$ for 
different bottom hadrons.  We now discuss these observable quantities in detail 
below.

\subsection {Input Parameters}

The basic parameters involved  are $m_c$, $\mu$, $\kappa_1$ and $\kappa_2$.
Using eq.(\ref{massl1l2}) and the measured masses of the ground state heavy 
mesons and heavy baryons, one can find that 
$\kappa_1$ in the heavy baryon system is almost
the same as the one in the heavy meson system. Thus besides the known 
masses\cite{pdg}
\begin{eqnarray}
m_{B^0}=5.2792~\mbox{GeV}; \quad m_{B^{*0}}=5.3249~\mbox{GeV};  \nonumber  \\
m_{D^+}=1.8693~\mbox{GeV}; \quad m_{D^{*+}}=2.0100~\mbox{GeV};  	  \\
m_{\Lambda_c}=2.2849~\mbox{GeV}; \quad m_{\Lambda_b}=5.624~\mbox{GeV} ; 
 \quad m_\tau=1.777~\mbox{GeV} ,\nonumber
\end{eqnarray}
there are only four parameters $m_c$, $\mu$, $\kappa_1$ and $\kappa_2$ 
in our calculations. Here $\kappa_1$ could be computed by QCD sum 
rules\cite{detl1} or other phenomenological model\cite{cheung}. It may also 
be extracted from fitting the meson spectra. Nevertheless, all the results 
remain suffering from large uncertainties. Here we shall use the most 
conservative range
\begin{eqnarray}
0.3~\mbox{GeV}^2 \leq -\kappa_1 \leq 0.7~\mbox{GeV}^2.
\end{eqnarray}
For $\kappa_2$, the value extracted from the
known $B-B^*$ mass splitting is quite stable
\begin{eqnarray}
\kappa_2 = {1 \over 8} (m^2_{B^{*0}}-m^2_{B^0}) =0.06~\mbox{GeV}^2 ,
\end{eqnarray}
which is accurate up to power correction of 
 $\bar{\Lambda}/2m_b  \sim 5\% $. The mass parameter
$m_c$ arising from the propagator in eqs.(\ref{qm}) and (\ref{gm})
shall be taken as an effective pole mass. Its value has
a large range\cite{pdg} 
$$1.2~\mbox{GeV} \leq m^{pole}_c \leq 1.9~\mbox{GeV}.$$
Note that $m_c^{pole}$ is different from the `heavy quark mass' in the
Lagrangian eq.(\ref{Leff}) and the `dressed heavy quark' mass in
eq.(\ref{addmass}). In the numerical calculations, we will restrict its range 
to be $$1.55~\mbox{GeV} \leq m^{pole}_c \leq 1.75~\mbox{GeV}.$$

\subsection{Lifetime Ratios}

  It is seen that, in the framework of new formulation of HQEFT, only 
`dressed bottom quark' mass $\hat{m}_b$ enters into the decay rate, but 
both charm quark mass $m_c$ and `dressed charm quark' mass $\hat{m}_c$
appear in the phase space factors at tree level.
So the lifetime ratio is not merely given by the ratio
$[m_b+\bar{\Lambda}(H)]^5/[m_b+\bar{\Lambda}(H^\prime)]^5$.
The $\mu$, $m_c$ and $\kappa_1$ dependences of the lifetime ratios
$\tau(B^0_s)/\tau(B^0)$ and $\tau(\Lambda_b)/\tau(B^0)$ are plotted in Fig. 1. 
It is seen that the ratios are not sensitive to the energy scale $\mu$. 
For a large range of parameters $m_c$ and $\kappa_{1}$, the ratios only 
slightly change, but for large $\kappa_{1}$, the ratios become very sensitive 
to a large charm quark mass. The ratios as functions of $m_c$ and $\kappa_{1}$ 
also exhibit some minimal points around which they change slowly. When taking 
$1.55~\mbox{GeV} \leq m_c \leq 1.75~\mbox{GeV}$ and
$-0.7~\mbox{GeV}^2 \leq \kappa_1 \leq -0.3~\mbox{GeV}^2$, we have
\begin{mathletters}
\begin{eqnarray}
{\tau(B^0_s) \over \tau(B^0)} &=& 0.94 \pm 0.04,  \\
{\tau(\Lambda_b) \over \tau(B^0)} &=& 0.76 \pm 0.06 ,
\end{eqnarray}
\end{mathletters}
which show a good agreement with the experimental data\cite{exratio}.

\subsection {Numerical Results in $B^0$ Decays}

  There is a puzzle in $b$ decays in the usual HQET that the predicted 
semileptonic branching ratio is significantly greater than the experimental
data. A large QCD enhancement of $b \to c\bar{c}s$ was expected to
suppress the value of the ratio, but it will lead to a much larger
charm counting than the world average.

  The theoretical predictions for $B_{SL}$ and $B_\tau$ strongly depend on 
the energy scale $\mu$, as shown in Fig. 2 and Fig. 3.
At $\mu=m_b$ and $\mu=m_b/2$, we have
\begin{mathletters}
\begin{eqnarray}
   B_{SL} &=& \cases{
    11.98\pm 0.60\%  ;& $\mu=m_b$, \cr
    10.00\pm 0.53\%  ;& $\mu=m_b/2$, \cr} \\
   \phantom{ \bigg[ }
   B_\tau  &=& \cases{
    3.23\mp 0.21\%  ;& $\mu=m_b$, \cr
    2.69\mp 0.19\%  ;& $\mu=m_b/2$. \cr}
\end{eqnarray}
\end{mathletters}
Their uncertainties caused by the charm quark mass and $\kappa_{1}$ are 
anti-correlated. One may compare it with the usual HQET prediction as shown 
in Fig. 6a and 6b.

  But their ratio $R$ is not sensitive to the choice of the scale $\mu$ as 
shown in Fig. 4. As the function of $m_{c}$ and $\kappa_{1}$, it shows a 
minimal point around which its variations become slow.  For $1.55~\mbox{GeV} 
\leq m_c \leq 1.75~\mbox{GeV}$ and
$-0.7~\mbox{GeV}^2 \leq \kappa_1 \leq -0.3~\mbox{GeV}^2$, we have
\begin{eqnarray}
 0.24 \leq R \leq 0.32 ,
\end{eqnarray}
here the main uncertainties arise from the uncertainties of $\kappa_1$ and 
$m_c$. One may also compare it with the usual HQET prediction as shown in 
Fig. 6c. For $m_b = 4.7 \pm 0.1$ GeV, $a=0.29 \pm 0.03$ and 
$\lambda_{1} = -0.36~\mbox{GeV}^{2}$, 
one has $ 0.18 < R < 0.26$. This shows one of the differences 
between the framework of new formulation of HQEFT and the usual HQET. 
The CLEO data for $B_{SL}$ and $B_\tau$ is\cite{cleo}
\begin{mathletters}
\begin{eqnarray}
B_{SL} = 10.5 \pm 0.5 \%, \\
B_\tau = 2.6 \pm 0.1 \%,
\end{eqnarray}
\end{mathletters}
which leads to $R= 0.25 \pm 0.02$. 

  In analogous to $R$, $n_c$ is not sensitive to $\mu$, but it strongly 
depends on $m_c$ and $\kappa_{1}$. The uncertainty in the theoretical 
prediction mainly arises from that of the value $m_c$ as shown in Fig. 5. 
 For $1.45~\mbox{GeV} \leq m_c \leq 1.80~\mbox{GeV}$ and $-0.7~\mbox{GeV}^2 
\leq \kappa_1 \leq -0.3~\mbox{GeV}^2$,  the numerical result for $n_c$ is
\begin{eqnarray}
n_c = 1.18 \pm 0.06 \pm 0.03 .
\end{eqnarray}
The uncertainties arising from the scale $\mu$ and $\Lambda_{QCD}$ are not 
more than 0.03. Note that its $m_c$ dependence shows a minimal point around 
which its variation is relatively slow. One may compare it with the usual HQET 
prediction as shown in Fig. 6c. For $\mu = m_{b}/2 - 2m_{b}$, 
$m_b = 4.7 \pm 0.1$ GeV, $a=0.29 \pm 0.03$ 
and $\lambda_{1} = -0.36~\mbox{GeV}^{2}$, one has $n_c = 1.22-1.34$.

  From Figs. 2-5, one can see that there exists a common
region for parameters $\mu$, $m_c$ and $\kappa_1$, so that all the quantities
$B_{SL}$, $B_\tau$, $R$ and $n_c$
are consistent with experimental data. When taking the following
interesting region for the parameters
\begin{eqnarray}
1.55~\mbox{GeV} \leq m_c \leq 1.75~\mbox{GeV} ,  \nonumber  \\
2.4~\mbox{GeV} \leq \mu \leq 4.4~\mbox{GeV}, \\
-0.7~\mbox{GeV}^2 \leq \kappa_1 \leq -0.3~\mbox{GeV}^2, \nonumber
\end{eqnarray}
 and normalizing $B_{SL}= 10.48 \pm 0.50\%$ and $R=0.26 \pm 0.02$,  we have
\begin{eqnarray}
n_c &=& 1.22 \mp 0.05.
\end{eqnarray}
which is consistent with the world average value. Note that a larger 
$m_c$ value leads to a lower $R$ and $n_c$ but to a larger $B_{SL}$ 
and $B_\tau$.

  In the usual HQET, up to $1/m^2_Q$ order and neglecting the `spectator effect',
however, there does not exist a common region for any
choice of parameters, which makes all the quantities to be consistent with the 
experimental data. This can been explicitly seen from Fig. 6.,  
where we have plotted four quantities $B_{SL}$, $B_\tau$, $R$ and 
$n_c$ as the functions of the running energy scale $\mu$ and the mass ratio 
$a \equiv m_c/m_b$ with $m_b=4.8\pm 0.2~\mbox{GeV}$ (Note that in the usual 
HQET, the $b$ quark mass $m_b$, instead of the `dressed heavy quark' mass 
$\hat{m}_b$, is the basic parameter). It is clear from Fig. 6 that there are 
two obstacles in the usual HQET to obtain a consistent fit. Firstly, a lower 
$n_c$ value requires a larger mass ratio $a$ and scale $\mu$, while a higher 
scale will lead semileptonic branching ratio to be larger than the experimental 
bound, and a higher mass ratio will result in a lower value $R$.

\subsection{Predictions for $B^0_s$ and $\Lambda_b$ Decays}

  We now discuss more about $B^0_s$ and $\Lambda_b$ decay. In the usual HQET,
all the above observable quantities in the $B^0$, $B^0_s$ and $\Lambda_b$ 
decays get the same results up to $1/m^2_b$ order. This is because the 
semileptonic decay rates of all the bottom hadrons are the same, thus the usual 
HQET leads to the following predictions up to $1/m^2_b$ order
\begin{mathletters}
\begin{eqnarray}
{B_{SL}(B^0_s) \over B_{SL}(B^0)} &=& {\Gamma_{B^0}^t \over \Gamma^t_{B^0_s}}
               =\left({\tau(B^0_s) \over \tau(B^0)}\right)_{th} \approx 1,\\
{B_{SL}(B^-) \over B_{SL}(B^0)} &=& {\Gamma_{B^0}^t \over \Gamma^t_{B^-}}
               =\left({\tau(B^-) \over \tau(B^0)}\right)_{th} \approx 1,\\ 
{B_{SL}(\Lambda_b) \over B_{SL}(B^0)} &=& 
     {\Gamma_{B^0}^t \over \Gamma^t_{\Lambda_b}}
              = \left({\tau(\Lambda_b) \over \tau(B^0)}\right)_{th} \approx 1.
\end{eqnarray}
\end{mathletters}
In the usual HQET, the lifetime ratio problems are expected to be solved 
through the non-leptonic decays. If it is the case, one should have the
following consequences for the ratios of the semileptonic branching ratios
\begin{mathletters}
\label{uhqet}
\begin{eqnarray}
{B_{SL}(B^0_s) \over B_{SL}(B^0)} &=& 
     \left({\tau(B^0_s) \over \tau(B^0)}\right)_{exp}=0.94\pm 0.04,\\
{B_{SL}(B^-) \over B_{SL}(B^0)} &=& 
     \left({\tau(B^-) \over \tau(B^0)}\right)_{exp}=1.07\pm 0.03,\\     
{B_{SL}(\Lambda_b) \over B_{SL}(B^0)} 
               &=& \left({\tau(\Lambda_b) \over \tau(B^0)}\right)_{exp}
               =0.79\pm 0.05.
\end{eqnarray}
\end{mathletters}
For the ratio $R$ one has
\begin{eqnarray}
\label{r}
R(B^0) \approx R(B^-) \approx R(B_s^0) \approx R(\Lambda_b).
\end{eqnarray}
One may compare it with the current experimental data\cite{pdg}
\begin{mathletters}
\begin{eqnarray}
B_{SL}(B^0) =10.48\pm 0.5\% ,\\
B_{SL}(B^-)=10.18\pm 0.5\% ,  \\
B(B^0_s \to D^- \ell^+ \bar{\nu}~\mbox{anything})=8.1\pm 2.5\%,   \\
B(\Lambda_b \to \Lambda_c \ell \bar{\nu}~\mbox{anything})=9.0^{+3.0}_{-2.8}\% .
\end{eqnarray}
\end{mathletters}

  In the framework of new formulation of HQEFT, as we have discussed in the 
previous sections, the picture for heavy hadron containing a single heavy quark 
is such that the heavy quark in the hadron is off-shell, its off-shellness is 
characterized by the binding energy $\bar{\Lambda}$. 
Consequently, the so-called `dressed heavy quark' masses $\hat{m}_b$ and 
$\hat{m}_c$ enter into the phase space factors. For $B^0_s$ and 
$\Lambda_b$ decays, the differences of the phase space factors already appear 
at tree level. It is not difficult to obtain the following relations 
\begin{mathletters}
\label{hqeft}
\begin{eqnarray}
& & B_{SL}(B^0) < B_{SL}(B^0_s) < B_{SL}(\Lambda_b), \\
& & R(B^0) \alt R(B^0_s) < R(\Lambda_b),  \\
& & n_{c}(B^0) > n_{c}(B^0_s) > n_{c}(\Lambda_b) .
\end{eqnarray}  
\end{mathletters}

 Compared eqs.(\ref{uhqet}) and eq.(\ref{r}) with eq.(\ref{hqeft}), one may find 
the significant difference between the usual HQET and the new formulation. 
The more and more accurate experiments in the future will provide results of 
${\cal B}_{SL}$ and $R$ to test the usual HQET and the new formulation of HQEFT. 

  When normalizing $B_{SL}(B^0)$ to $9.98\% \sim 10.98\%$ and $R(B^0)$ to
$0.24 \sim 0.28$, we have
\begin{mathletters}
\begin{eqnarray}
B_{SL}(B^0_s)=10.75 \pm 0.67\% ;&& \quad B_{SL}(\Lambda_b)=11.09 \pm 0.91\% ,\\
{B_{SL}(B^0_s) \over B_{SL}(B^0)} = 1.02 \pm 0.02 ;&\ \ \ & \quad 
{B_{SL}(\Lambda_b) \over B_{SL}(B^0)} = 1.06 \pm 0.05 ,\\
R(B^0_s) = 0.28 \mp 0.02 ;&& \quad R(\Lambda_b) = 0.32 \mp 0.02,\\
n_c(B^0_s) = 1.20 \pm 0.04 ;&& \quad n_c(\Lambda_b) = 1.18 \pm 0.05.
\end{eqnarray}
\end{mathletters}

\subsection{$|V_{cb}|$ from Inclusive Decays}

  One of the most important applications is the determination of the CKM matrix 
element $|V_{cb}|$. From eq.(\ref{sl}) the semileptonic decay rate is given by 
\begin{eqnarray}
\label{vsl}
\Gamma(b\to c e \bar{\nu}) &=& {G^2_F \hat{m}^5_b V^2_{cb} \over 192 \pi^3} 
  \eta_{ce}(\rho,0,\mu) 
    	\{ I_0(\rho,0,\hr)+I_1(\rho,0,\hr) {A \over \hat{m}^2_b} + 
    	I_2(\rho,0,\hr) {N_b \over \hat{m}^2_b} + \cdots \} ,
\end{eqnarray}
where the ellipse denotes higher order perturbative and non-perturbative 
corrections. It is explicitly seen that only the combination of $\bar{\Lambda}$ 
and $m_b$, i.e., $\hat{m}_b$, appears in the above equation and there is
no $1/m_b$ corrections. At the same time, the $1/m_b$ power corrections to
the decay width $\Gamma(b\to c e \bar{\nu})$  is as small as $-0.7\sim 5\% $.
Thus all the parameters in eq.(\ref{vsl}) have clear physical meaning and the 
higher power corrections can be ignored. This allows us to extract $|V_{cb}|$ 
from fitting the experimental data. The value of $|V_{cb}|$ as function of
$m_c$ and $\mu$ is shown in Fig. 7. It is easy to find that the value of 
$|V_{cb}|$ has only a weak dependence on the energy scale $\mu$ and
the charm quark mass $m_c$ and $\kappa_{1}$ cause the main uncertainties for 
the prediction. There exist maximal points around which its variation 
as the function of $m_c$ and $\kappa_{1}$ becomes slow. The values of 
$|V_{cb}|$ around those points should be more reliable.  It is 
interesting to note that once one normalizes the semileptonic branching ratio 
to the experimental data, $|V_{cb}|$ exhibits an interesting correlation with
the ratio $R$. Its relation is shown in Fig. 7d. It is seen that the value of 
$|V_{cb}|$ increases linearly as $R$ decreases. Including two-loop QCD 
corrections and fitting  $Br(b\to c e \bar{\nu})$ to the measured data, 
we obtain
\begin{eqnarray}
|V_{cb}|=0.0388 \pm 0.0005_{exp} \pm 0.0012_{th} , 
\end{eqnarray}
where the theoretical uncertainties arise from those of
$R = 0.25 \pm 0.03$ and $\kappa_1=-0.5 \pm 0.2~\mbox{GeV}^2$.
The experimental uncertainty comes from the errors of $B^0$ 
lifetime. One may compare the above prediction for $|V_{cb}|$ in the new 
formulation of HQEFT with the one in the usual HQET which is plotted in Fig. 8.

\section{Conclusions}

  We have presented, within the framework of new formulation of HQEFT, a general 
formulation for the inclusive decays of bottom hadrons via the heavy quark 
expansion. Such a general formulation has exhibited interesting features:
The $1/m_{b}$ corrections are absent and the mass preliminarily entered into 
the general formulation is neither the bottom quark mass $m_b$ nor heavy hadron 
mass $m_{H}$, it is the so-called `dressed bottom quark' mass $\hat{m}_{b}$ that is 
actually the heavy hadron mass in the infinite quark mass limit, i.e., 
$\hat{m}_b = \lim_{m_b \to \infty} m_{H_b} = m_{b} + \bar{\Lambda} 
= m_{H_b}[1 + O(1/m^2_{H_b})]$. Consequently, the decay rates of the bottom 
hadrons can be reexpressed in terms of the bottom hadron mass with the 
$1/m_{b}$ corrections remaining absent. This feature is now consistent with 
the expectation from Luke's theorem, since according to that theorem the 
sum of all the channels of exclusive $b$ decays is free from the $1/m_b$ 
order corrections at zero recoil. Such a feature distinguishes from the one 
obtained by Chay-Georgi-Grinstein theorem within the framework of the usual 
HQET. The reason is that in the new formulation of HQEFT, Luke's theorem comes 
out automatically without the need of imposing the equation of motion 
$iv\cdot D Q_{v}^{(+)}=0$. As a consequence, not only the lifetime ratio 
$\tau(\Lambda_b)/\tau(B^0)$ but also the ratio $\tau(B^0_s)/\tau(B^0)$ is in 
good agreement with the experimental data.  At the same time, the results for 
the inclusive semileptonic branching ratio $B_{SL}$, the ratio $R$ and the 
charm counting $n_c$ are also consistent with the experimental data. In 
particular, the CKM matrix element $|V_{cb}|$ has been nicely extracted from 
the inclusive semileptonic decay rate, and the result well agrees with the one 
from the exclusive decays\cite{WWY}. For $m_c = 1.75~\mbox{GeV}$ and 
$\kappa_1 = -0.4~\mbox{GeV}^2$, we have
\begin{mathletters}
\begin{eqnarray}
{\tau(B^0_s) \over \tau(B^0)} = 0.95; &&\quad 
  {\tau(\Lambda_b) \over \tau(B^0)} = 0.77,\\
B_{SL}(B^0_s) = 10.82\%; &&\quad B_{SL}(\Lambda_b) = 11.32\%,\\
R(B^0) = 0.25; &\quad R(B^0_s) = 0.27; &\quad R(\Lambda_b) = 0.32,\\
n_c(B^0) = 1.19; &\quad n_c(B^0_s) = 1.17; &\quad n_c(\Lambda_b) = 1.13 , \\
|V_{cb}| = 0.0398.
\end{eqnarray}
\end{mathletters}
If the nonspectator effects are taken into account, the charm counting will 
decrease further\cite{Neubert}. It is expected that more precise data for 
the $B^0$ decays and further test for the predictions on the $B^0_s$ and 
$\Lambda_b$ systems will provide a useful check for the framework of  
new formulation of HQEFT at the level of higher order corrections. 
  
\acknowledgments

We would like to thank professor Y.B. Dai for useful discussions. 
This work was supported in part by the NSF of China under the grant No. 19625514.

\ \\
\begin{center}
Table. 1. The observable parameters in $B^0$ and $B^0_s$ as well as $\Lambda_b$
decays with $\mu=m_b$.
\end{center}
\begin{center}
\begin{tabular}{|c|c|c|c|c|c|c|c|c|c|c|c|c|} \hline
$m_c$(GeV)   & \multicolumn{3}{|c|}{1.55 } 
   & \multicolumn{3}{|c|}{ 1.65 }
   & \multicolumn{3}{|c|}{ 1.75 }
   & \multicolumn{3}{|c|}{ 1.8 }   \\  \cline{1-13}
$\kappa_1(~\mbox{GeV}^2)$   
   & $-0.5$   & $-0.6$ & $-0.7$  &
     $-0.4$   & $-0.5$ & $-0.6$ &
     $-0.3$   & $-0.4$ & $-0.5$ &
     $-0.2$   & $-0.3$ & $-0.4$  \\ \hline
${\tau(B^0_s) \over \tau(B^0)}$
  & 0.903 & 0.915 & 0.946  & 0.913& 0.931 & 0.969  & 0.923 & 0.945  
  & 0.987 & 0.922  & 0.938 & 0.970  \\ \hline
${\tau(\Lambda_b) \over \tau(B^0)}$
  & 0.701 & 0.703 & 0.724 & 0.718  & 0.729 & 0.764  & 0.738 & 0.757 
   & 0.804 & 0.743  & 0.757 & 0.793  \\ \hline
$B_{SL}(B^0)(\%)$
  & 11.62 & 11.56 & 11.46 & 12.04  & 11.92 & 11.74  & 12.41 & 12.20  
  & 11.94 & 12.68  & 12.44 & 12.16  \\ \hline
$B_\tau(B^0)$
  & 0.032 & 0.031 & 0.030 & 0.032  & 0.031 & 0.030  & 0.031 & 0.031  & 0.030 
  & 0.032  & 0.031 & 0.030  \\ \hline
${\cal R}(B^0)$
  & 0.27 & 0.26 & 0.26 & 0.26 & 0.26  & 0.26 & 0.25 & 0.25  & 0.25 
  & 0.25  & 0.25 & 0.25  \\ \hline
$n_c(B^0)$
  & 1.21 & 1.22 & 1.23 & 1.18 & 1.19  & 1.21 & 1.16 & 1.17  & 1.19 
  & 1.14  & 1.16 & 1.18  \\ \hline
$B_{SL}(B^0_s)(\%)$
  & 11.68 & 11.66 & 11.63 & 12.19 & 12.12  & 12.04 & 12.68 & 12.55  
  & 12.39 & 12.99  & 12.82 & 12.63  \\ \hline
$B_\tau(B^0_s)$
  & 0.034 & 0.033 & 0.032 & 0.034 & 0.033  & 0.032 & 0.034 & 0.033  & 0.032 
  & 0.035  & 0.034 & 0.032  \\ \hline
${\cal R}(B^0_s)$
  & 0.29 & 0.28 & 0.27 & 0.28 & 0.27  & 0.27 & 0.27 & 0.26  & 0.26 
  & 0.27  & 0.26 & 0.26  \\ \hline
$n_c(B^0_s)$
  & 1.21 & 1.21 & 1.21 & 1.17 & 1.18  & 1.18 & 1.14 & 1.15  & 1.16 
  & 1.12  & 1.13 & 1.14  \\ \hline
${B_{SL}(B^0_s) \over B_{SL}(B^0)}$
  & 1.00 & 1.01 & 1.02 & 1.01 & 1.02  & 1.03 & 1.02 & 1.03  & 1.04 
  & 1.02  & 1.03 & 1.04  \\ \hline
$B_{SL}(\Lambda_b)(\%)$
  & 11.73 & 11.72 & 11.73 & 12.37 & 12.35  & 12.33 & 13.08 & 13.02  & 12.96 
  & 13.49  & 13.40 & 13.32  \\ \hline
$B_\tau(\Lambda_b)$
  & 0.039 & 0.038 & 0.037 & 0.040 & 0.039  & 0.038 & 0.041 & 0.040  & 0.039 
  & 0.042  & 0.041 & 0.040  \\ \hline
${\cal R}(\Lambda_b)$
  & 0.33 & 0.32 & 0.32 & 0.32 & 0.32  & 0.31 & 0.31 & 0.31  & 0.30 
  & 0.31  & 0.30 & 0.30  \\ \hline
$n_c(\Lambda_b)$
  & 1.21 & 1.21 & 1.21 & 1.16 & 1.16  & 1.17 & 1.11 & 1.12  & 1.12 
  & 1.08  & 1.09 & 1.10  \\ \hline
${B_{SL}(\Lambda_b) \over B_{SL}(B^0)}$
  & 1.01 & 1.01 & 1.02 & 1.03 & 1.04  & 1.05 & 1.05 & 1.07  & 1.09 
  & 1.06  & 1.08 & 1.10  \\ \hline
$V_{cb}(10^{-2})$
  & 3.63 & 3.73 & 3.72 & 3.68 & 3.73  & 3.74 & 3.75 & 3.79  & 3.77 
  & 3.76  & 3.81 & 3.82  \\ \hline
\end{tabular}
\end{center}

\newpage
\begin{center}
Table. 2. The observable parameters in $B^0$ and $B^0_s$ as well as $\Lambda_b$
decays with $\mu=m_b/2$.
\end{center}
\begin{center}
\begin{tabular}{|c|c|c|c|c|c|c|c|c|c|c|c|c|} \hline
$m_c$(GeV)   & \multicolumn{3}{|c|}{1.55 } 
   & \multicolumn{3}{|c|}{ 1.65 }
   & \multicolumn{3}{|c|}{ 1.75 }
   & \multicolumn{3}{|c|}{ 1.8 }   \\  \cline{1-13}
$\kappa_1(~\mbox{GeV}^2)$   
   & $-0.5$   & $-0.6$ & $-0.7$  &
     $-0.4$   & $-0.5$ & $-0.6$ &
     $-0.3$   & $-0.4$ & $-0.5$ &
     $-0.2$   & $-0.3$ & $-0.4$  \\ \hline
${\tau(B^0_s) \over \tau(B^0)}$
  & 0.904 & 0.917 & 0.949  & 0.916& 0.934 & 0.973  & 0.927 & 0.949  
  & 0.993 & 0.927  & 0.943 & 0.976  \\ \hline
${\tau(\Lambda_b) \over \tau(B^0)}$
  & 0.706 & 0.708 & 0.731 & 0.725  & 0.737 & 0.774  & 0.748 & 0.769 
   & 0.818 & 0.755  & 0.770 & 0.809  \\ \hline
$B_{SL}(B^0)(\%)$
  & 9.53 & 9.47 & 9.38 & 9.93  & 9.81 & 9.64  & 10.28 & 10.08  
  & 9.83 & 10.53  & 10.30 & 10.03  \\ \hline
$B_\tau(B^0)$
  & 0.026 & 0.026 & 0.025 & 0.027  & 0.026 & 0.025  & 0.026 & 0.026  & 0.025 
  & 0.027  & 0.026 & 0.025  \\ \hline
${\cal R}(B^0)$
  & 0.28 & 0.27 & 0.27 & 0.27 & 0.26  & 0.26 & 0.26 & 0.25  & 0.26 
  & 0.26  & 0.25 & 0.25  \\ \hline
$n_c(B^0)$
  & 1.24 & 1.25 & 1.26 & 1.21 & 1.22  & 1.23 & 1.18 & 1.20  & 1.22 
  & 1.16  & 1.18 & 1.20  \\ \hline
$B_{SL}(B^0_s)(\%)$
  & 9.60 & 9.58 & 9.55 & 10.08 & 10.01  & 9.93 & 10.55 & 10.41  
  & 10.26 & 10.84  & 10.68 & 10.49  \\ \hline
$B_\tau(B^0_s)$
  & 0.028 & 0.027 & 0.027 & 0.029 & 0.028  & 0.027 & 0.029 & 0.028  & 0.027 
  & 0.030  & 0.028 & 0.027  \\ \hline
${\cal R}(B^0_s)$
  & 0.29 & 0.29 & 0.28 & 0.29 & 0.28  & 0.27 & 0.27 & 0.27  & 0.26 
  & 0.27  & 0.27 & 0.26  \\ \hline
$n_c(B^0_s)$
  & 1.24 & 1.24 & 1.24 & 1.20 & 1.20  & 1.21 & 1.16 & 1.17  & 1.19 
  & 1.14  & 1.15 & 1.17  \\ \hline
${B_{SL}(B^0_s) \over B_{SL}(B^0)}$
  & 1.01 & 1.01 & 1.02 & 1.02 & 1.02  & 1.03 & 1.03 & 1.03  & 1.04 
  & 1.03  & 1.04 & 1.05  \\ \hline
$B_{SL}(\Lambda_b)(\%)$
  & 9.69 & 9.68 & 9.68 & 10.30 & 10.27  & 10.25 & 10.97 & 10.91  & 10.85 
  & 11.38  & 11.29 & 11.20  \\ \hline
$B_\tau(\Lambda_b)$
  & 0.032 & 0.032 & 0.031 & 0.034 & 0.033  & 0.032 & 0.035 & 0.034  & 0.033 
  & 0.036  & 0.035 & 0.034  \\ \hline
${\cal R}(\Lambda_b)$
  & 0.34 & 0.33 & 0.32 & 0.33 & 0.32  & 0.32 & 0.32 & 0.31  & 0.30 
  & 0.31  & 0.31 & 0.30  \\ \hline
$n_c(\Lambda_b)$
  & 1.23 & 1.24 & 1.24 & 1.19 & 1.19  & 1.19 & 1.13 & 1.14  & 1.14 
  & 1.10  & 1.11 & 1.12  \\ \hline
${B_{SL}(\Lambda_b) \over B_{SL}(B^0)}$
  & 1.02 & 1.02 & 1.03 & 1.04 & 1.05  & 1.06 & 1.07 & 1.08  & 1.10 
  & 1.08  & 1.10 & 1.12  \\ \hline
$V_{cb}(10^{-2})$
  & 3.86 & 3.97 & 3.96 & 3.92 & 3.97  & 3.97 & 3.99 & 4.03  & 4.01 
  & 4.00  & 4.05 & 4.06  \\ \hline
\end{tabular}
\end{center}

\begin{center}
Table .3. The observable parameters in $B^0$ and $B^0_s$ as well as $\Lambda_b$
decays with $B_{SL} = 10.48\%$.
\end{center}
\begin{center}
\begin{tabular}{|c|c|c|c|c|c|c|c|c|c|c|c|c|} \hline
$m_c$(GeV)   & \multicolumn{3}{|c|}{1.55 } 
   & \multicolumn{3}{|c|}{ 1.65 }
   & \multicolumn{3}{|c|}{ 1.75 }
   & \multicolumn{3}{|c|}{ 1.8 }   \\  \cline{1-13}
$\kappa_1(~\mbox{GeV}^2)$   
   & $-0.5$   & $-0.6$ & $-0.7$  &
     $-0.4$   & $-0.5$ & $-0.6$ &
     $-0.3$   & $-0.4$ & $-0.5$ &
     $-0.2$   & $-0.3$ & $-0.4$  \\ \hline
$\mu(~\mbox{GeV})$
  & 2.85 & 2.91 & 3.00  & 2.53& 2.62 & 2.76  & 2.31 & 2.43  
  & 2.60 & 2.18  & 2.30 & 2.46  \\ \hline
${\tau(B^0_s) \over \tau(B^0)}$
  & 0.904 & 0.916 & 0.947  & 0.915& 0.933 & 0.971  & 0.927 & 0.949  
  & 0.991 & 0.927  & 0.942 & 0.975  \\ \hline
${\tau(\Lambda_b) \over \tau(B^0)}$
  & 0.704 & 0.706 & 0.727 & 0.723  & 0.735 & 0.770  & 0.747 & 0.766 
   & 0.814 & 0.755  & 0.769 & 0.805  \\ \hline
$B_\tau(B^0)$
  & 0.029 & 0.028 & 0.028 & 0.028  & 0.027 & 0.027  & 0.027 & 0.027  & 0.027 
  & 0.027  & 0.026 & 0.026  \\ \hline
${\cal R}(B^0)$
  & 0.27 & 0.27 & 0.26 & 0.27 & 0.26  & 0.26 & 0.26 & 0.25  & 0.26 
  & 0.26  & 0.25 & 0.25  \\ \hline
$n_c(B^0)$
  & 1.23 & 1.23 & 1.24 & 1.20 & 1.21  & 1.22 & 1.18 & 1.19  & 1.21 
  & 1.16  & 1.18 & 1.20  \\ \hline
$B_{SL}(B^0_s)(\%)$
  & 10.54 & 10.58 & 10.66 & 10.63 & 10.69  & 10.78 & 10.75 & 10.82  
  & 10.92 & 10.79  & 10.86 & 10.94  \\ \hline
$B_\tau(B^0_s)$
  & 0.031 & 0.030 & 0.029 & 0.030 & 0.029  & 0.029 & 0.029 & 0.029  & 0.028 
  & 0.029  & 0.029 & 0.028  \\ \hline
${\cal R}(B^0_s)$
  & 0.29 & 0.28 & 0.28 & 0.28 & 0.28  & 0.27 & 0.27 & 0.27  & 0.26 
  & 0.27  & 0.27 & 0.26  \\ \hline
$n_c(B^0_s)$
  & 1.22 & 1.23 & 1.23 & 1.19 & 1.20  & 1.20 & 1.16 & 1.17  & 1.18 
  & 1.14  & 1.15 & 1.16  \\ \hline
${B_{SL}(B^0_s) \over B_{SL}(B^0)}$
  & 1.01 & 1.01 & 1.02 & 1.01 & 1.02  & 1.03 & 1.03 & 1.03  & 1.04 
  & 1.03  & 1.04 & 1.04  \\ \hline
$B_{SL}(\Lambda_b)(\%)$
  & 10.62 & 10.67 & 10.77 & 10.85 & 10.94  & 11.09 & 11.18 & 11.32  & 11.51 
  & 11.33  & 11.47 & 11.65  \\ \hline
$B_\tau(\Lambda_b)$
  & 0.035 & 0.035 & 0.035 & 0.035 & 0.035  & 0.035 & 0.035 & 0.035  & 0.035 
  & 0.036  & 0.035 & 0.035  \\ \hline
${\cal R}(\Lambda_b)$
  & 0.33 & 0.33 & 0.32 & 0.32 & 0.32  & 0.31 & 0.32 & 0.31  & 0.30 
  & 0.31  & 0.31 & 0.30  \\ \hline
$n_c(\Lambda_b)$
  & 1.22 & 1.22 & 1.22 & 1.18 & 1.18  & 1.18 & 1.13 & 1.13  & 1.14 
  & 1.10  & 1.11 & 1.11  \\ \hline
${B_{SL}(\Lambda_b) \over B_{SL}(B^0)}$
  & 1.01 & 1.02 & 1.03 & 1.04 & 1.04  & 1.06 & 1.07 & 1.08  & 1.10 
  & 1.08  & 1.09 & 1.11  \\ \hline
$V_{cb}(10^{-2})$
  & 3.75 & 3.84 & 3.82 & 3.85 & 3.88  & 3.87 & 3.96 & 3.98  & 3.93 
  & 4.01  & 4.03 & 4.01  \\ \hline
\end{tabular}
\end{center}

\begin{figure}
\label{fig:tld}
\caption{The lifetime ratios $\tau(\Lambda_b)/\tau(B^0)$
and $\tau(B^0_s)/\tau(B^0)$ as the functions of the running energy scale $\mu$, 
charm quark mass $m_c$ and $\kappa_1$. (1a) and (1c) : $\mu=m_b$, 
(1b) and (1d) : $m_c=1.65$GeV.
The experimental data are $\tau(\Lambda_b)/\tau(B^0)=0.79 \pm 0.05$ and 
$\tau(B^0_s)/\tau(B^0)=0.94 \pm 0.04$.}
\end{figure}

\begin{figure}
\label{fig:bsl}
\caption{The semileptonic branching ratio $B_{SL}$ as the function of $\mu$, 
$m_c$ and $\kappa_1$. (2a) : $\mu=m_b/2$, (2b) : $m_c=1.65$GeV 
and (2c) : $\kappa_1=-0.6~\mbox{GeV}^2$.
The world average is $B_{SL}= 10.48 \pm 0.50\%$.}
\end{figure}

\begin{figure}
\label{fig:Bt}
\caption{$B_\tau$ as the function of $\mu$, $m_c$ and $\kappa_1$. 
(3a) : $\mu=m_b/2$, (3b) : $m_c=1.65$GeV and 
(3c) : $\kappa_1=-0.6~\mbox{GeV}^2$.
The CLEO data is $B_\tau=2.6 \pm 0.1\%$.}
\end{figure}

\begin{figure}
\label{fig:r}
\caption{$R$ as the function of $\mu$, $m_c$ and $\kappa_1$. 
(4a) : $\mu=m_b$, (4b) : $m_c=1.65$GeV and 
(4c) : $\kappa_1=-0.6~\mbox{GeV}^2$.
The CLEO data is $R=0.25 \pm 0.02$.}
\end{figure}

\begin{figure}
\label{fig:nc}
\caption{The charm counting $n_c$ as the function of $\mu$, 
$m_c$ and $\kappa_1$. (5a) : $\kappa_1=-0.6~\mbox{GeV}^2$, 
(5b) : $\mu=m_b$ and (5c) : $m_c=1.65$GeV, (d) : $B_{SL} = 10.48\%$.
The world average is $n_c= 1.17 \pm 0.04$.}
\end{figure}

\begin{figure}
\label{fig:pfig}
\caption{$B_{SL}$, $B_\tau$, $R$ and $n_c$ as functions of $\mu$, $m_b$ and
$a\equiv m_c/m_b$ in the usual HQET.
We have fixed $\lambda_1 = -0.36~\mbox{GeV}^2$. (6a), (6b) and (6d) :
$ m_b = 4.8 ~\mbox{GeV}$, (6c) : $\mu=4.8~\mbox{GeV}$.}
\end{figure}

\begin{figure}
\caption{$|V_{cb}|$ as functions of $\mu$, $m_c$ and $\kappa_1$. 
(7a) and (7c) : $\kappa_1=-0.6~\mbox{GeV}^2$, (7b) : $\mu=m_b/2$, 
(7d) : correlation between $|V_{cb}|$ and $R$.}
\end{figure}

\begin{figure}
\label{fig:pvcb}
\caption{$|V_{cb}|$ as function of $m_b$ and $a\equiv m_c/m_b$ in usual HQET.
With $B_{SL}=10.48\%$ and $\lambda_1 = -0.36~\mbox{GeV}^2$.}
\end{figure}



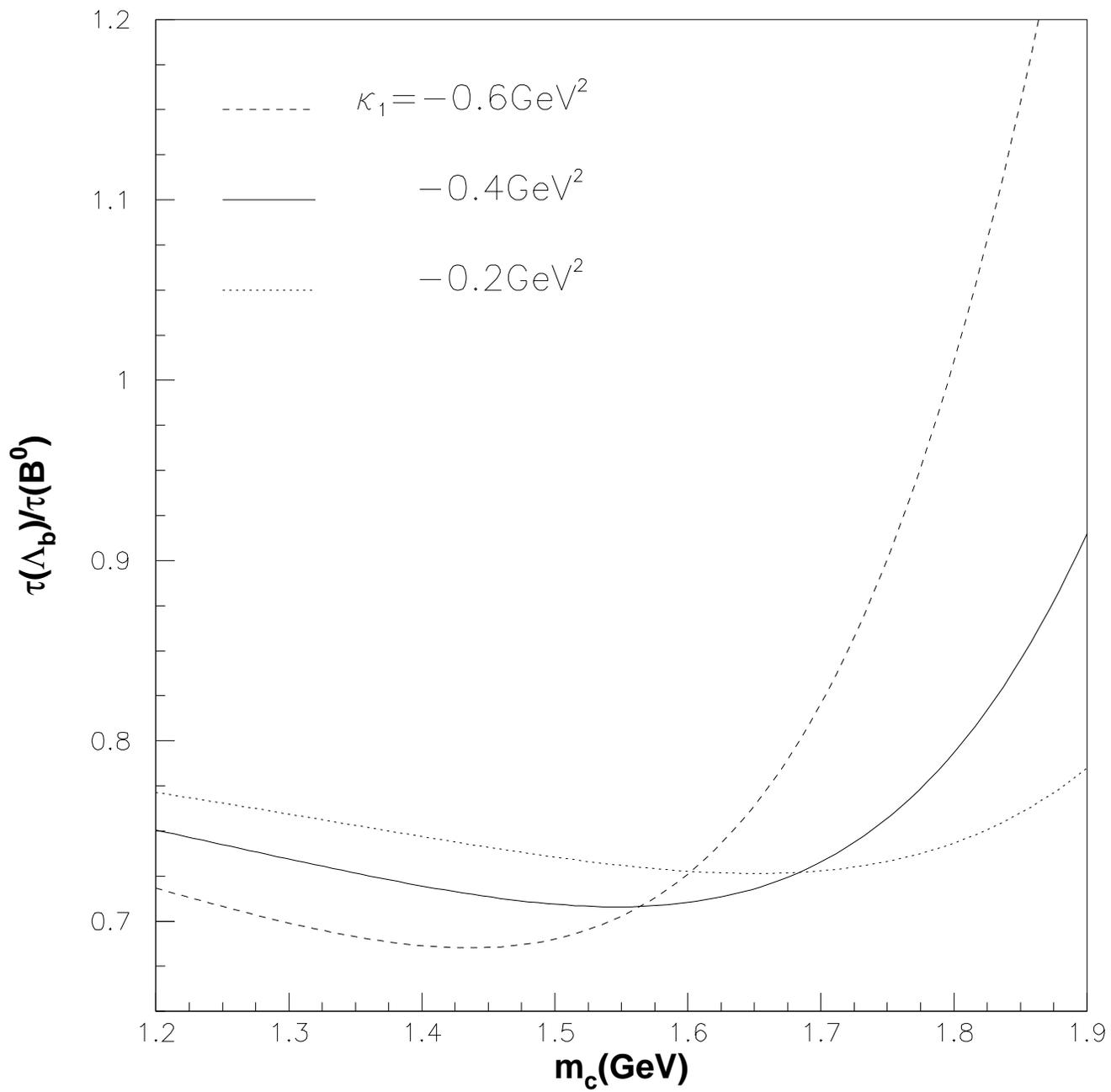
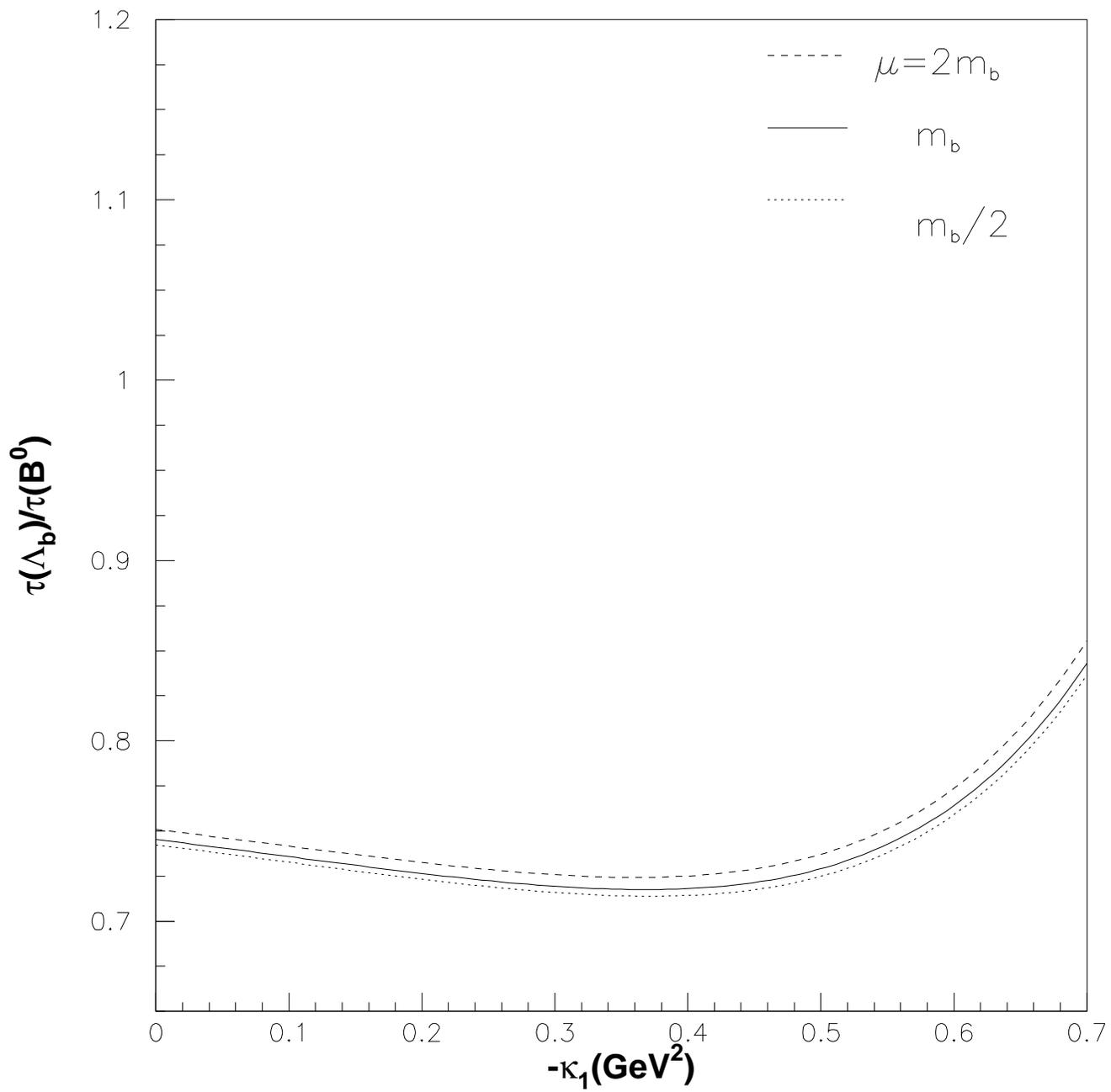
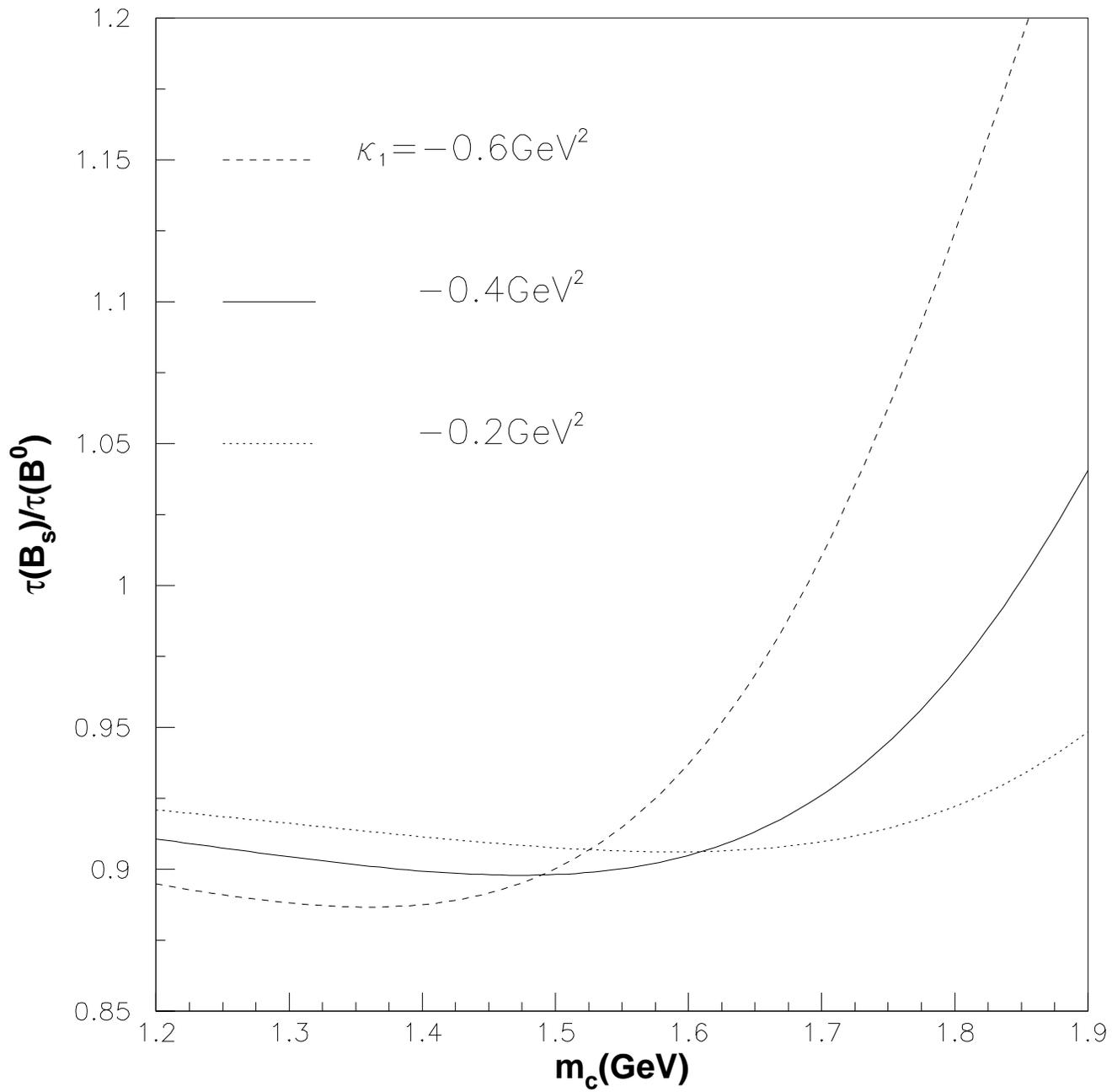
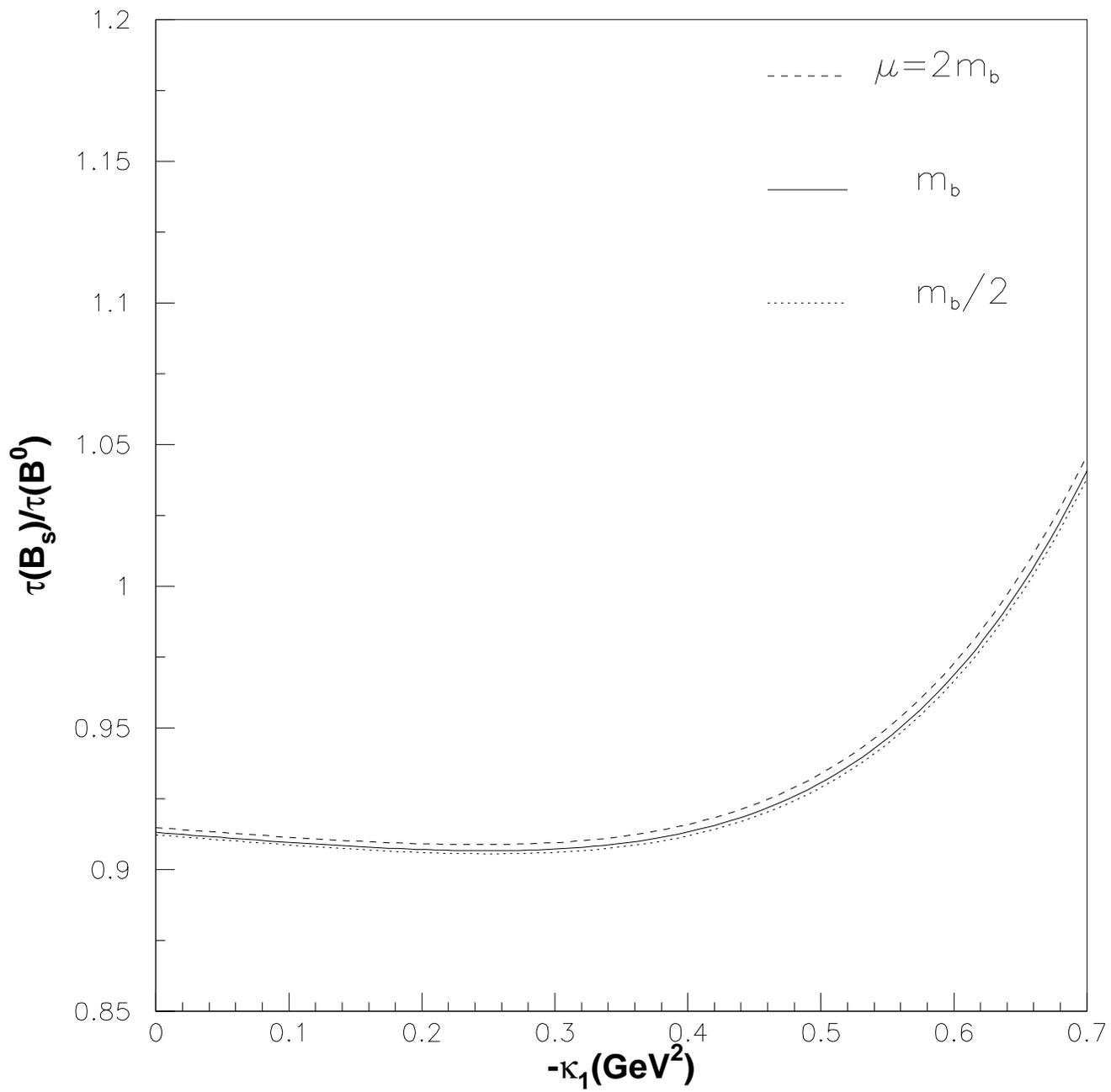
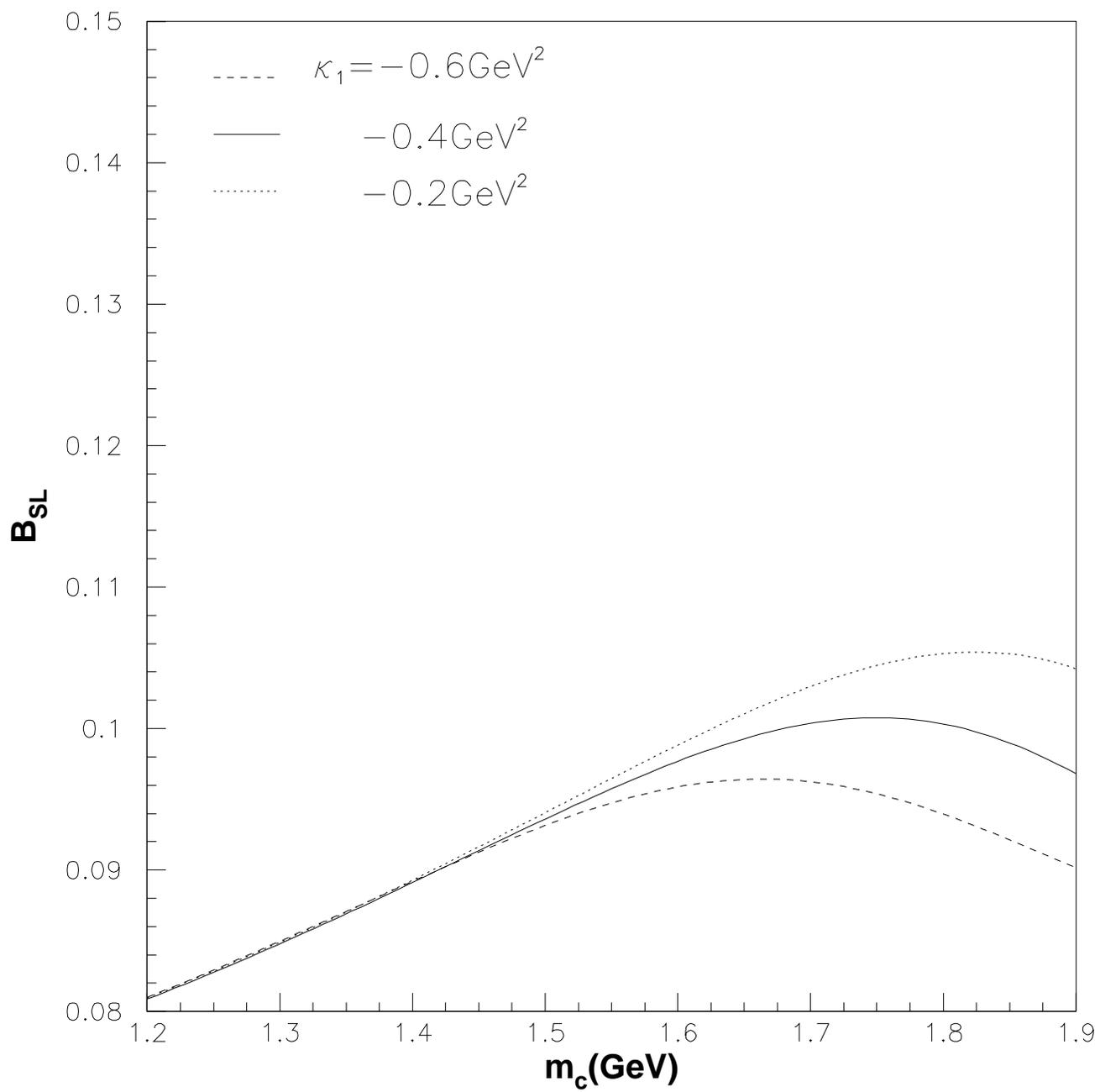
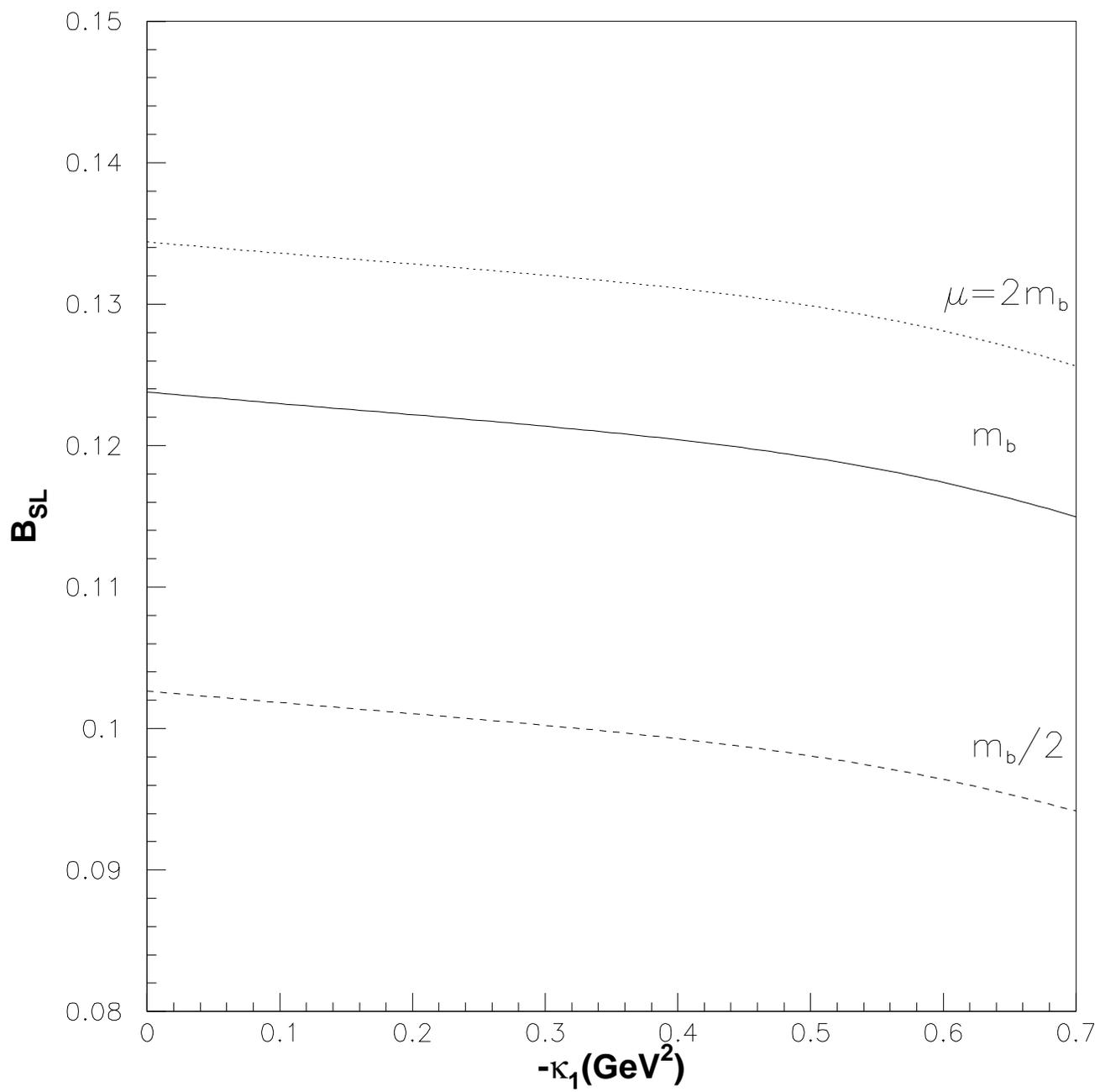
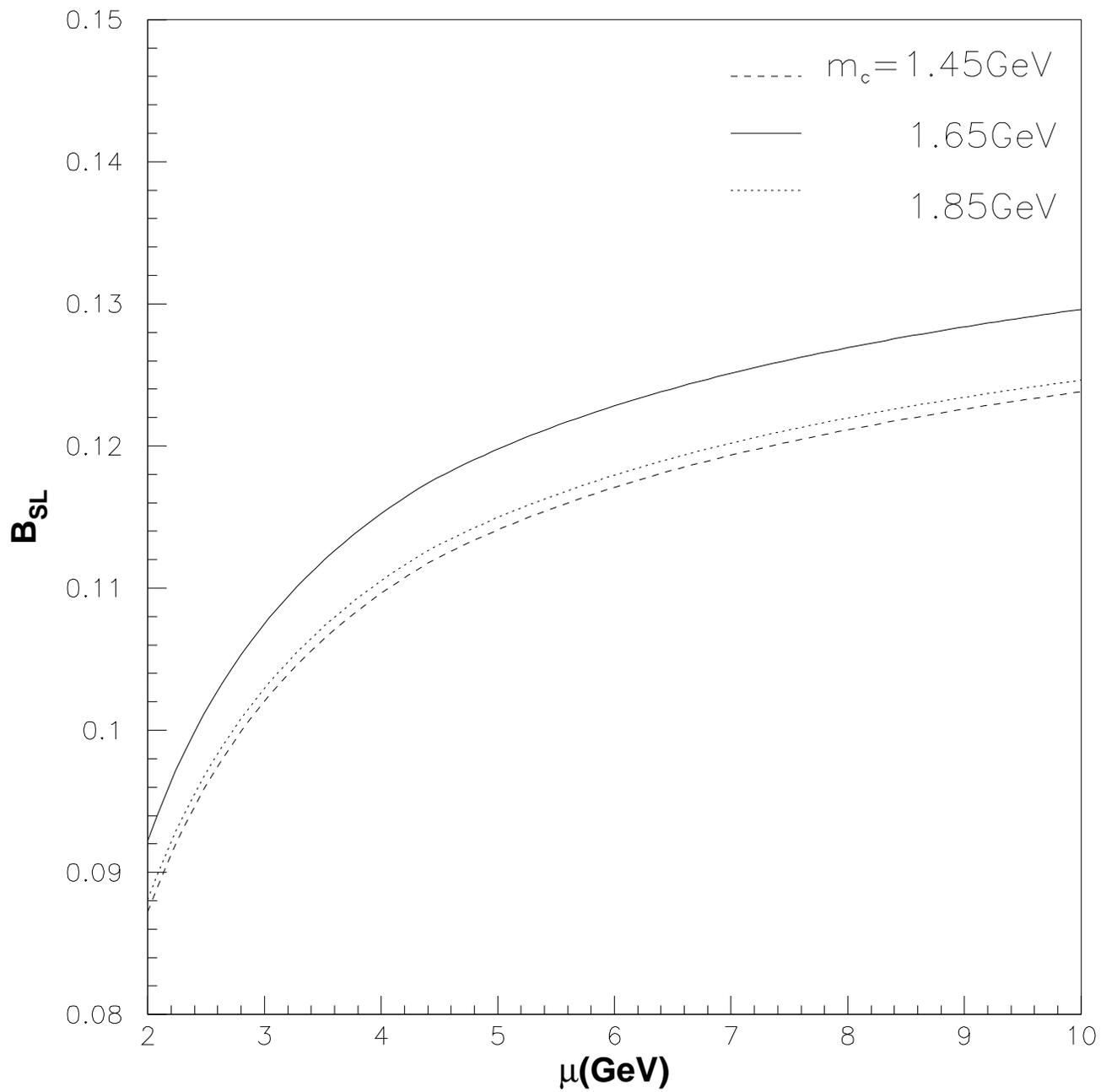
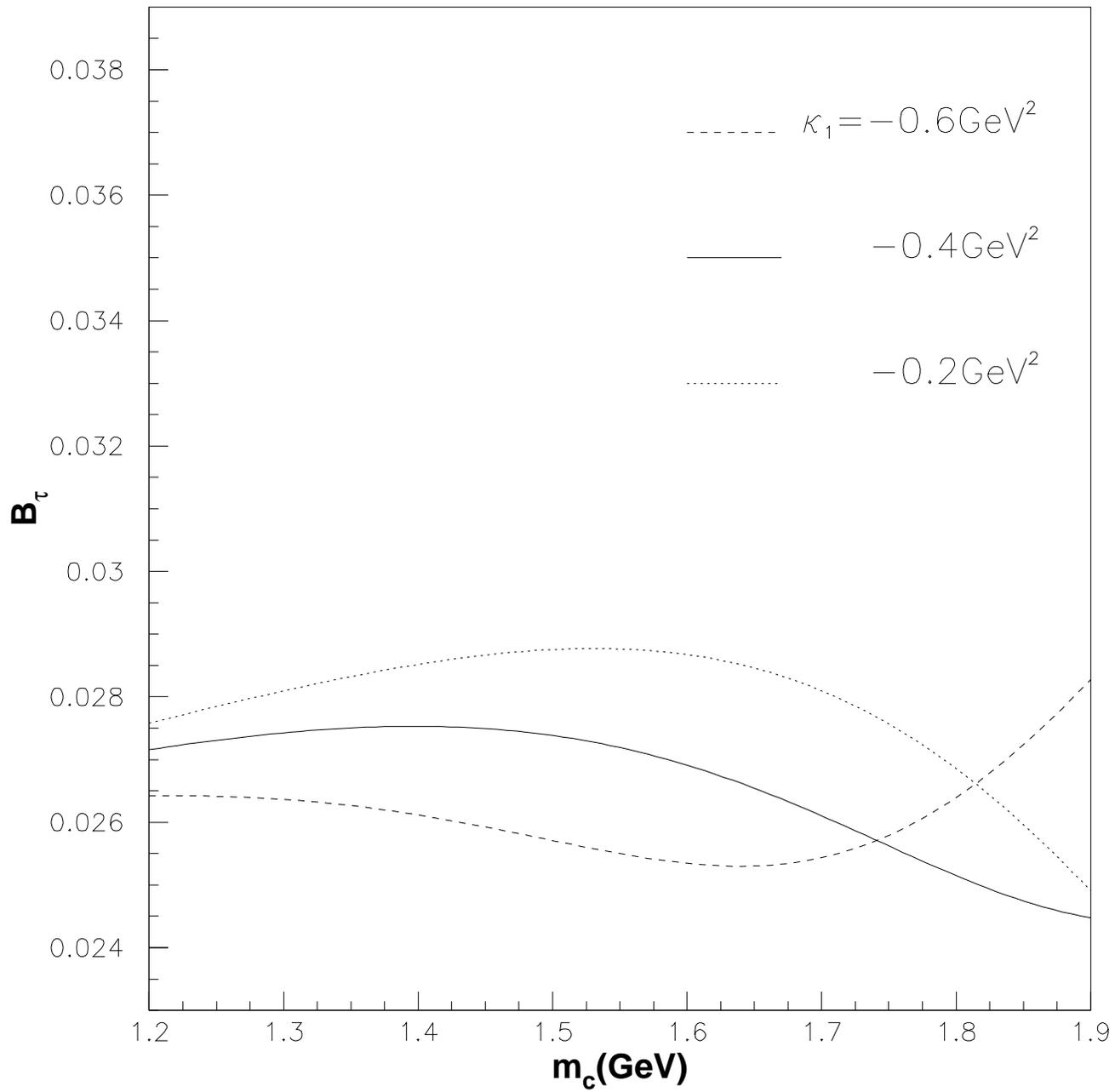
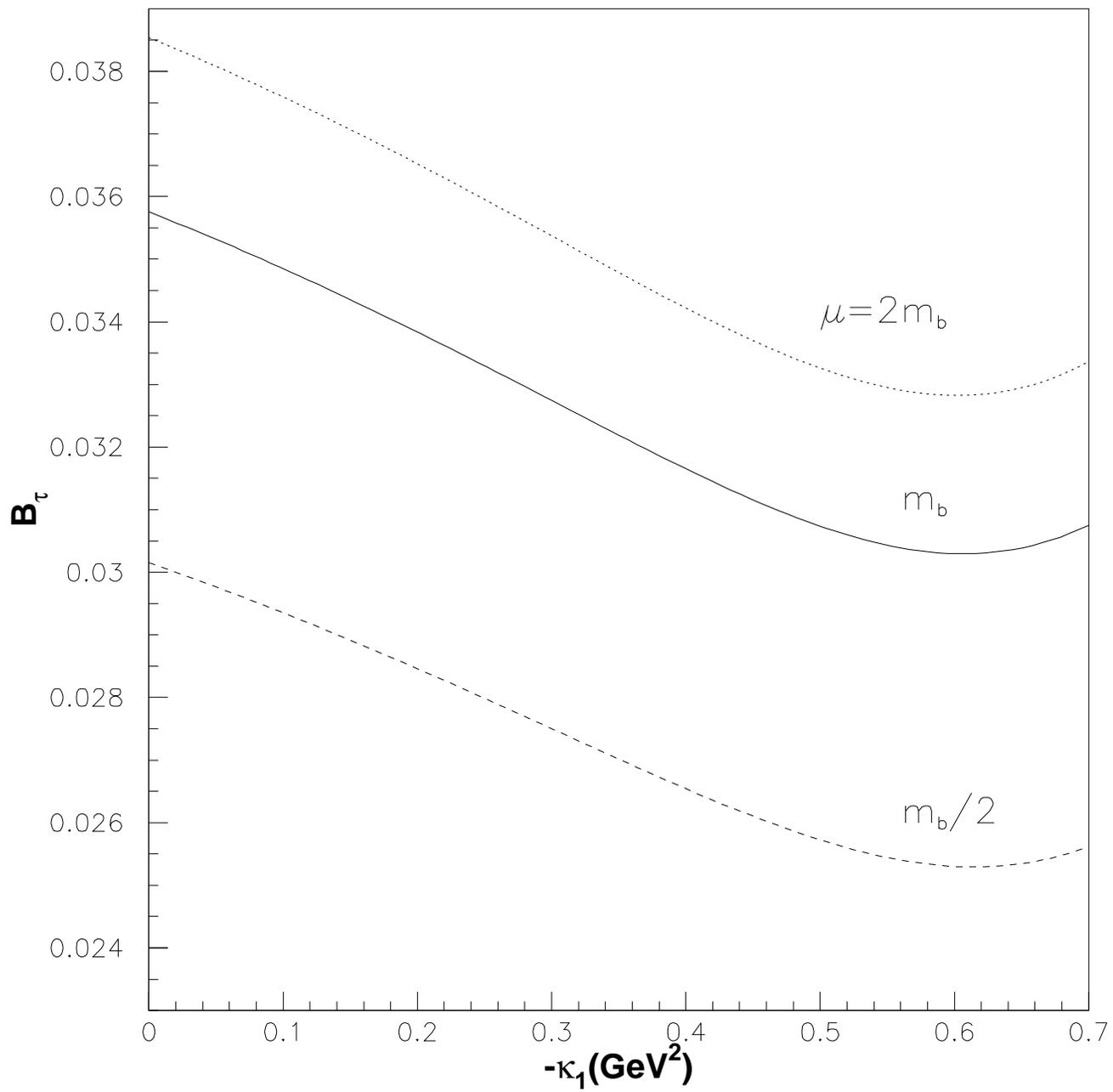
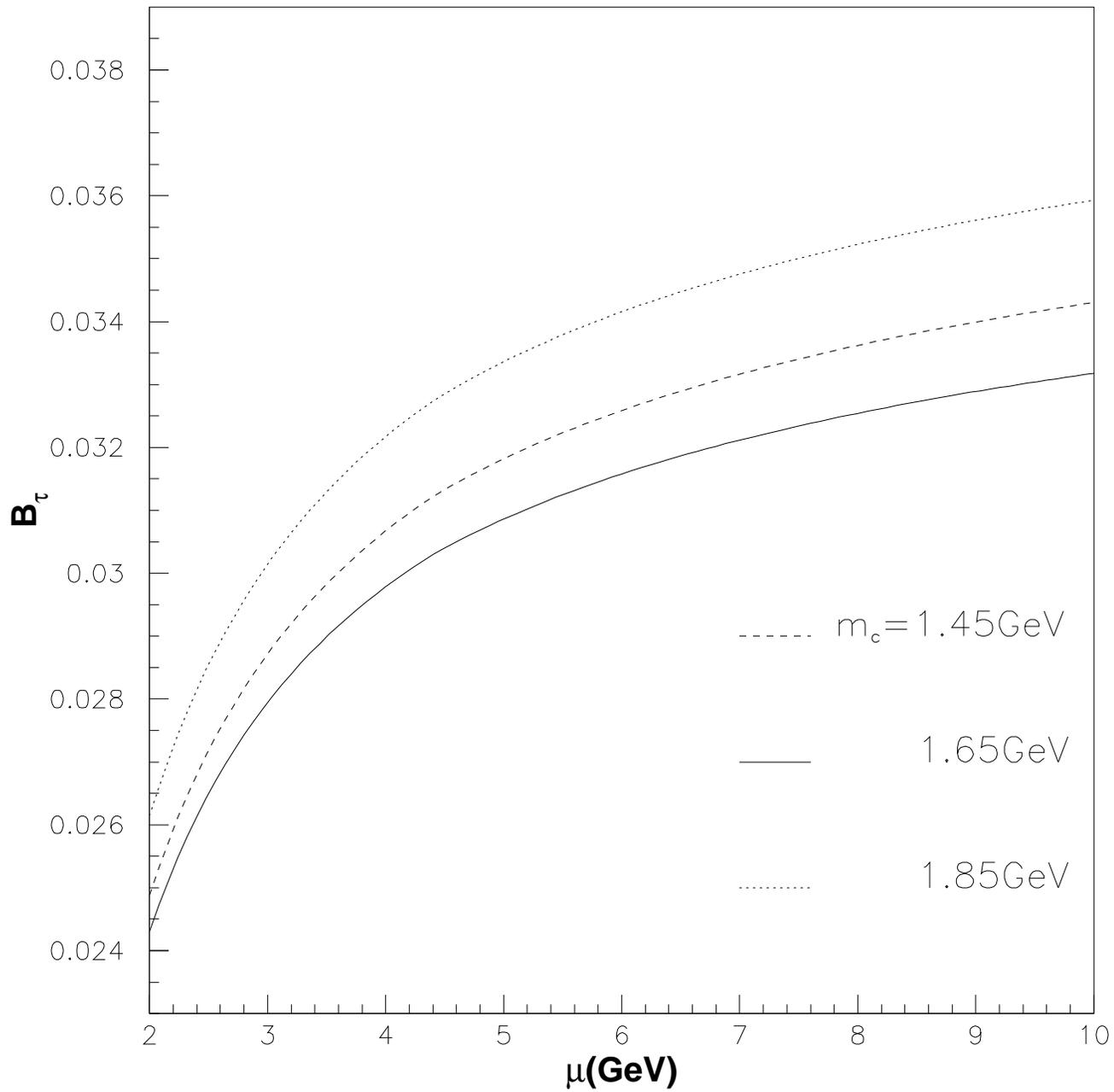
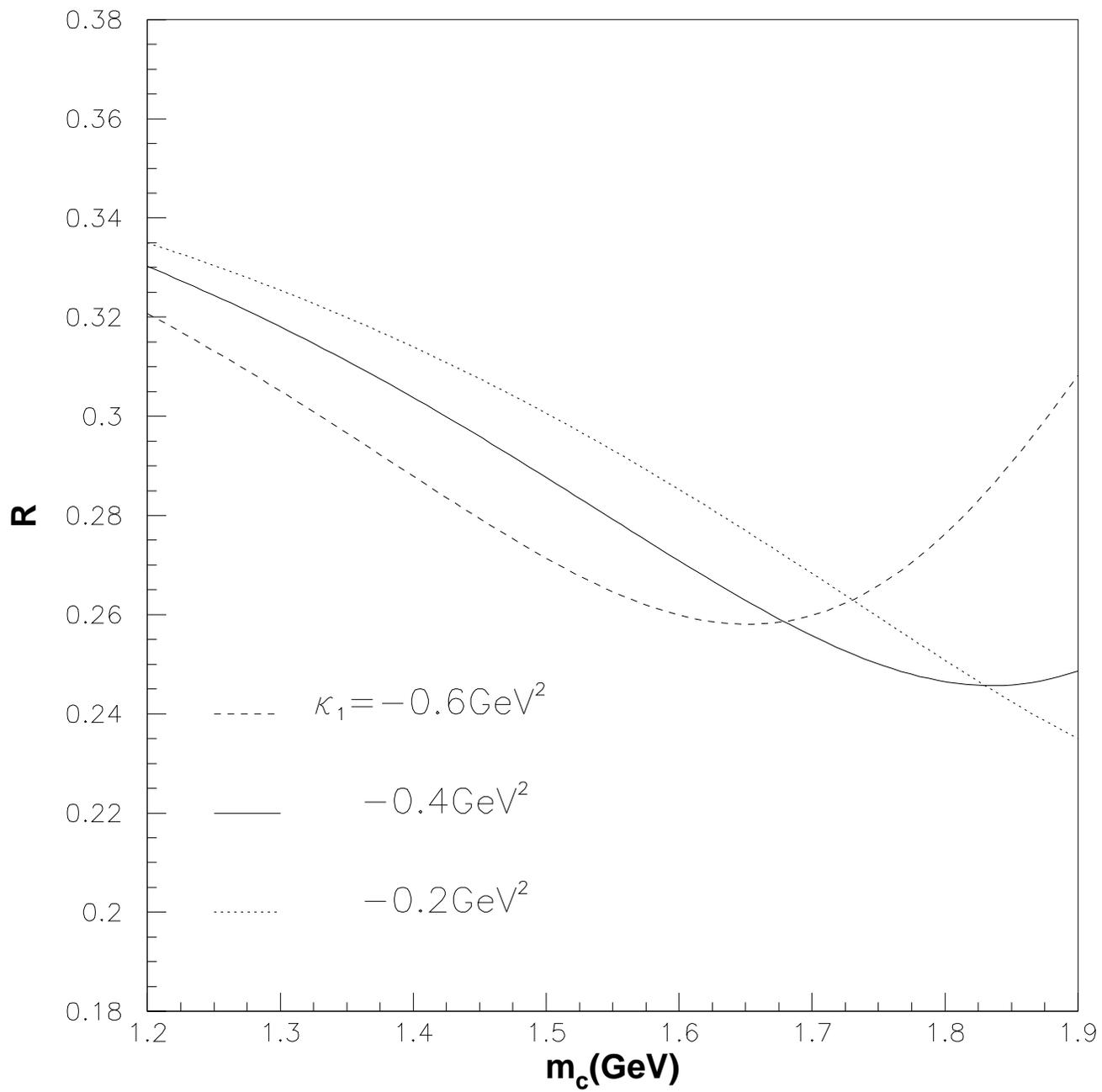
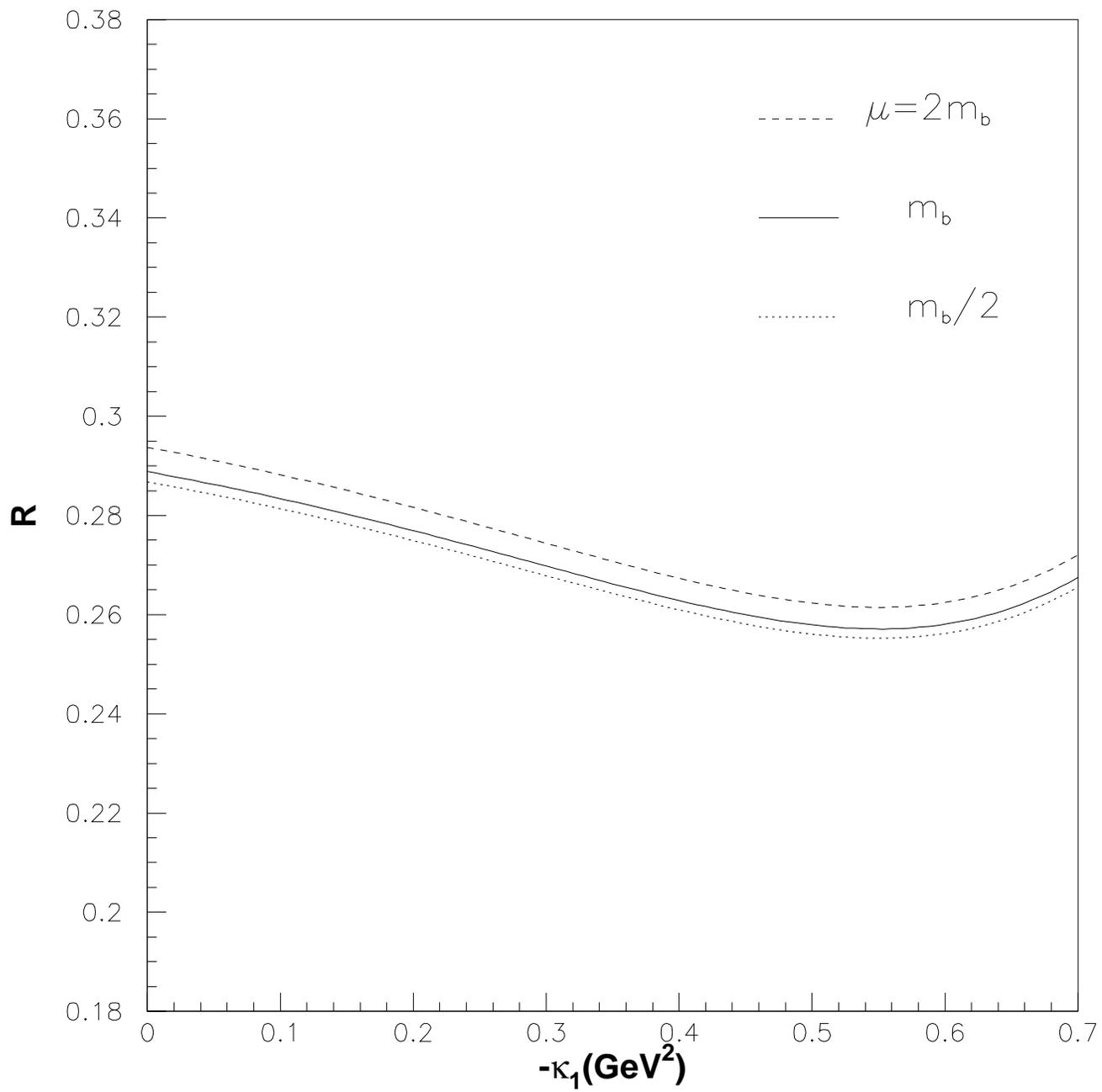
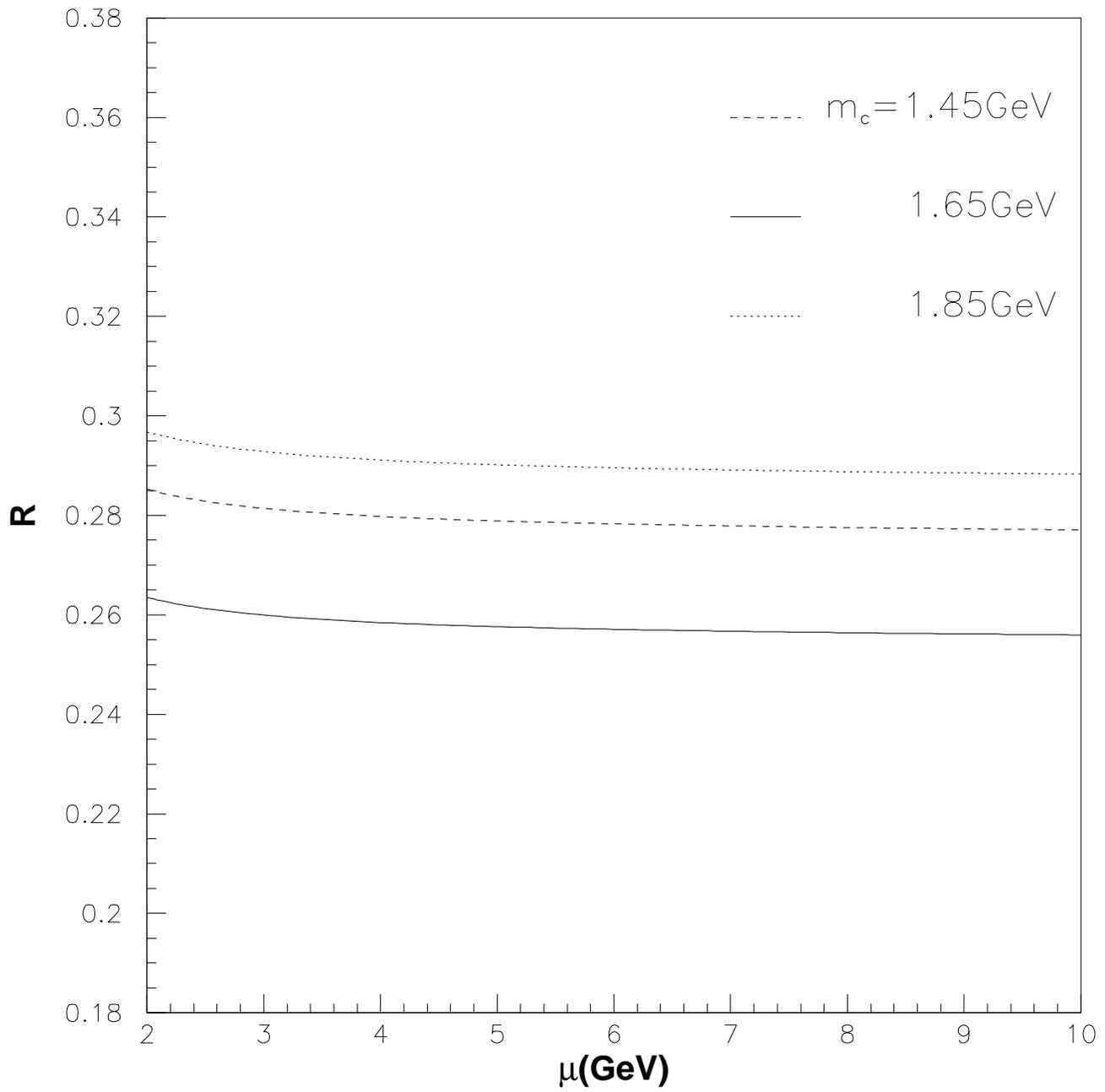
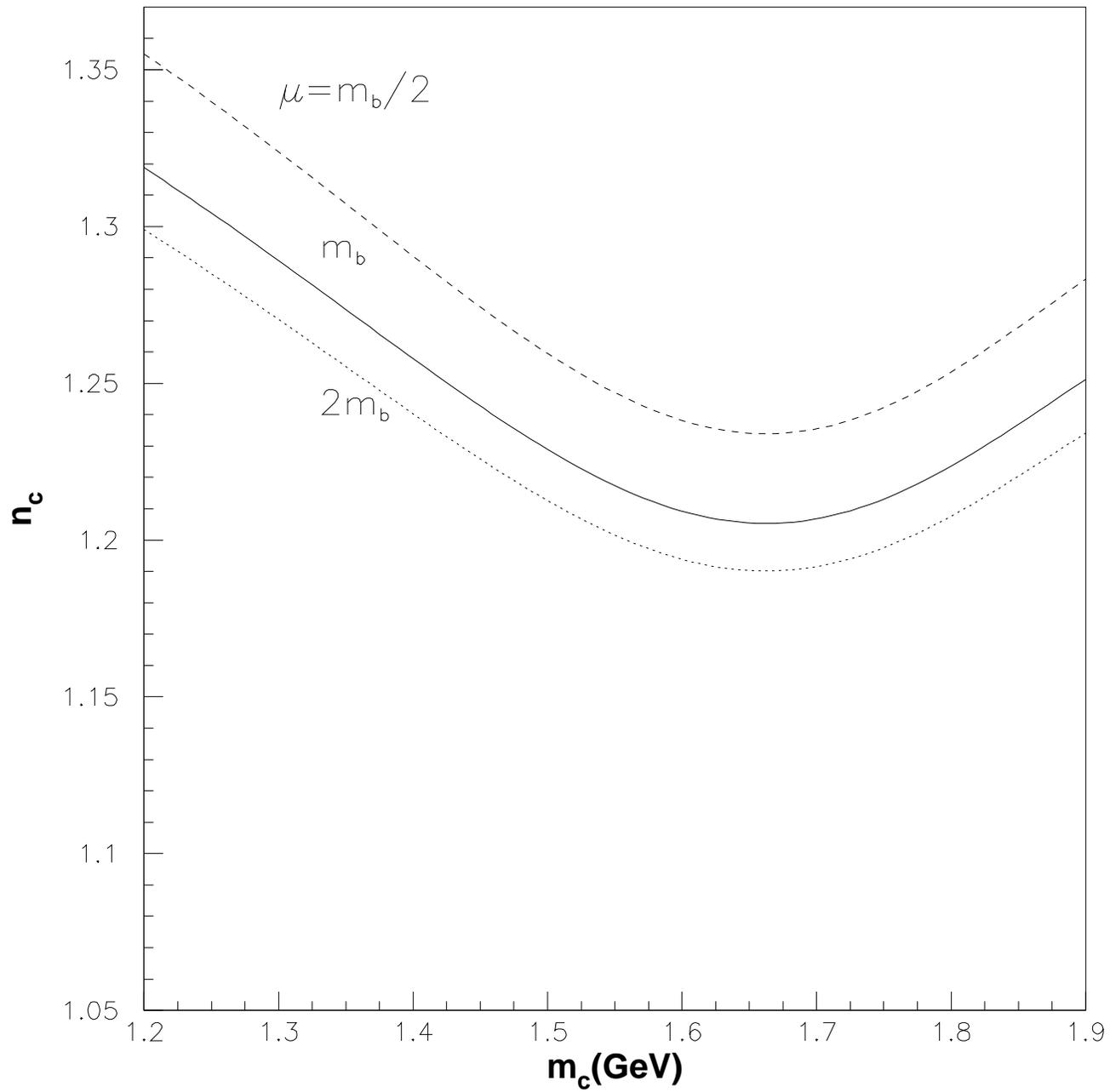
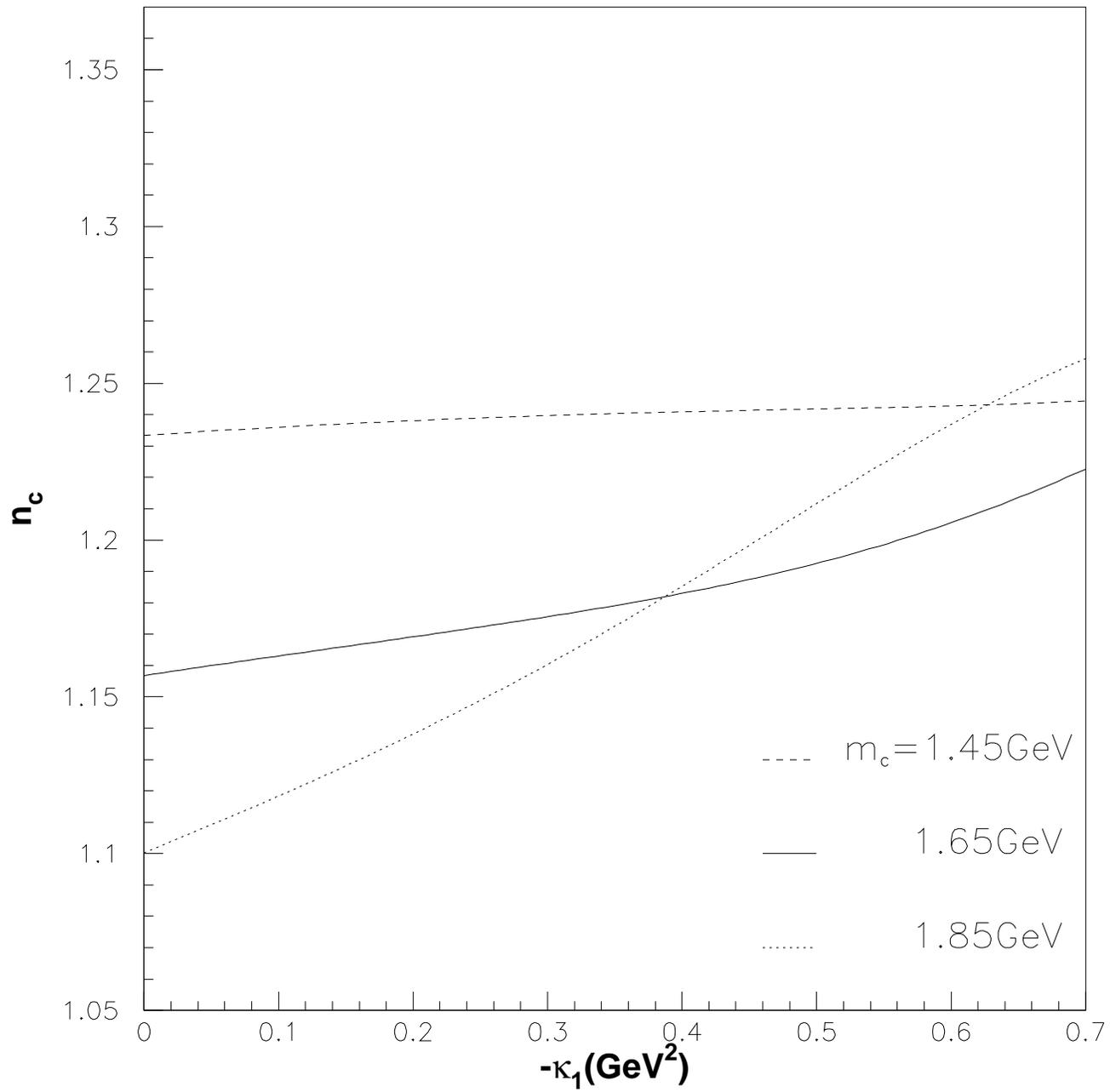
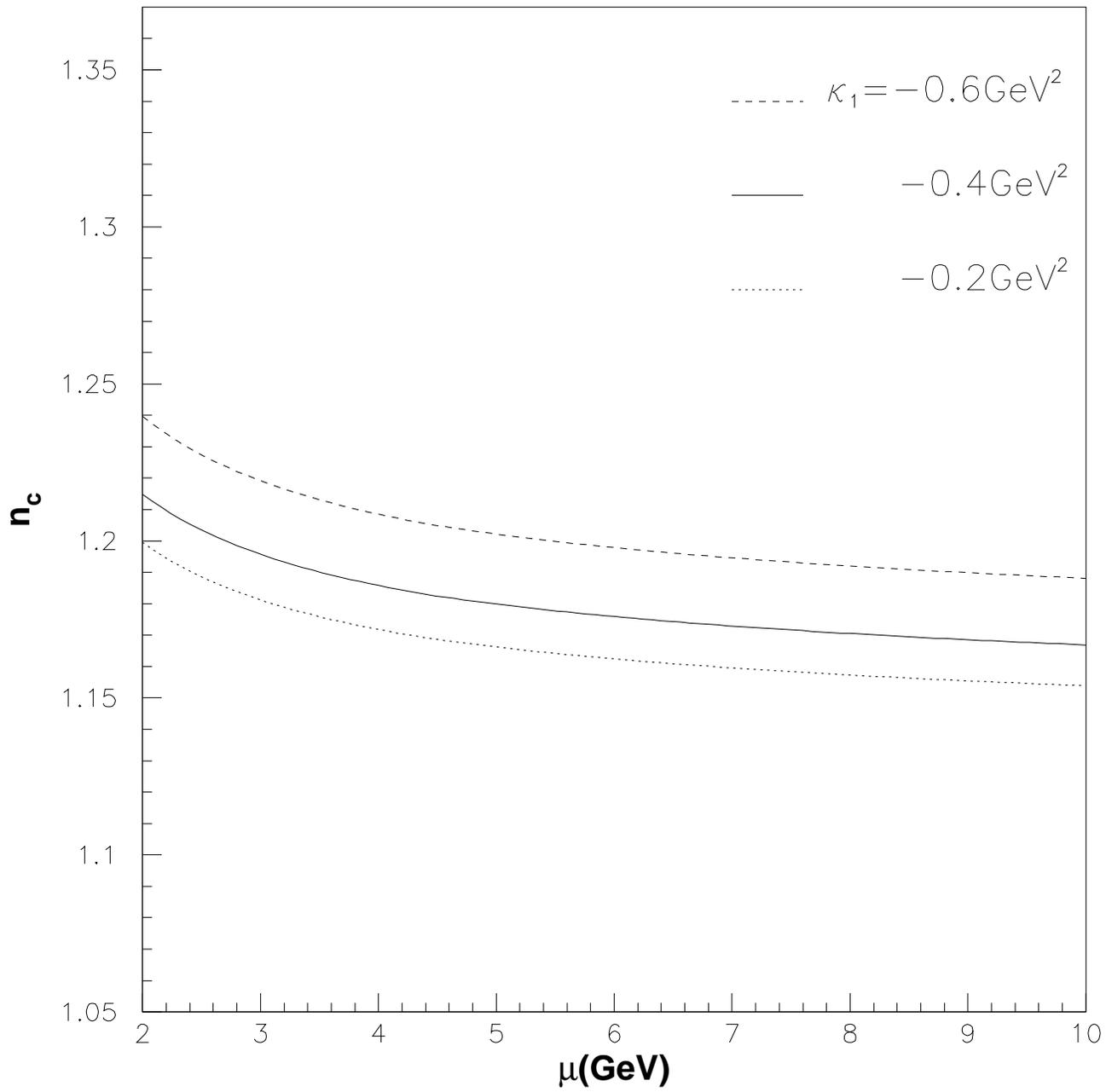
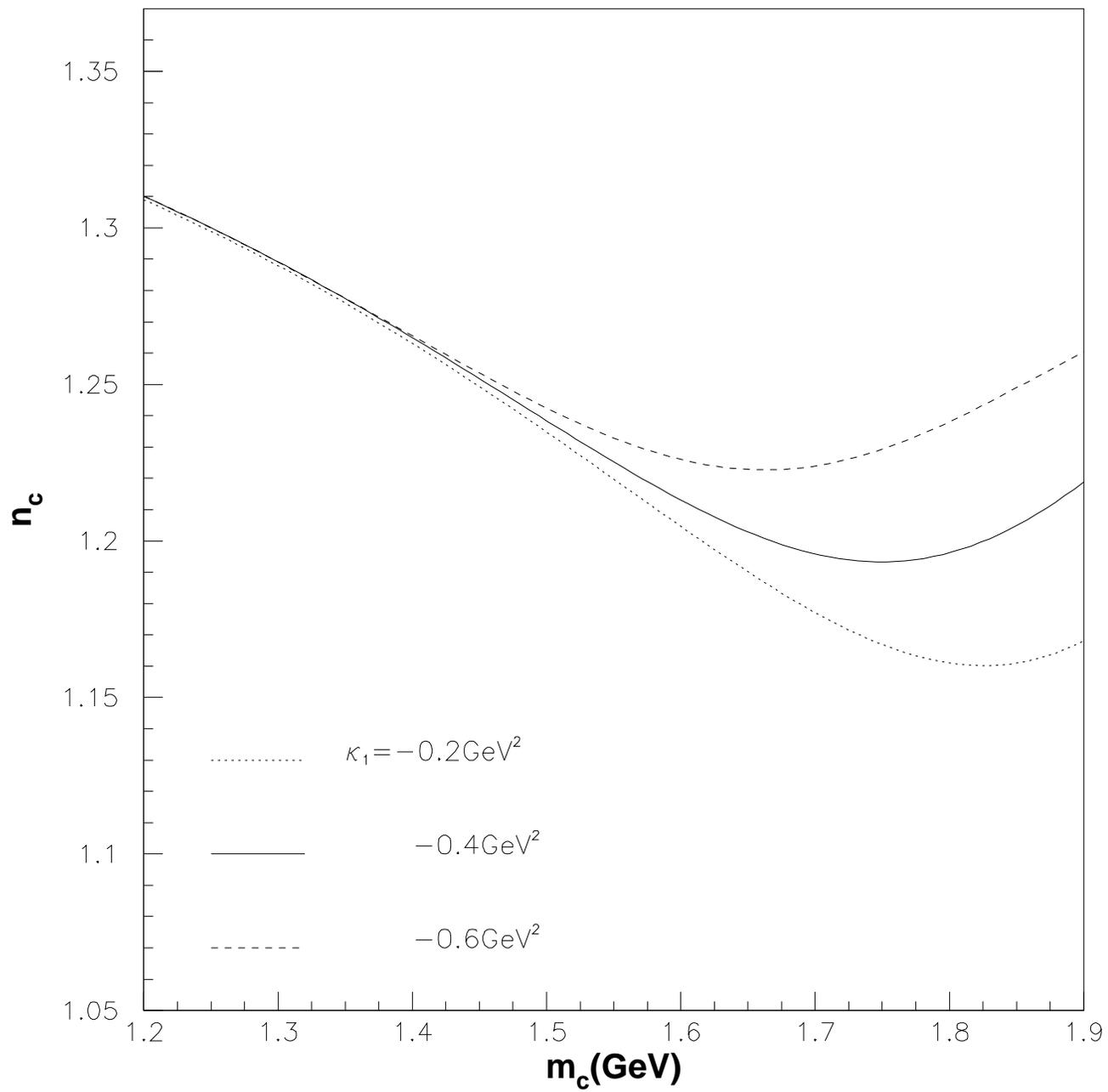
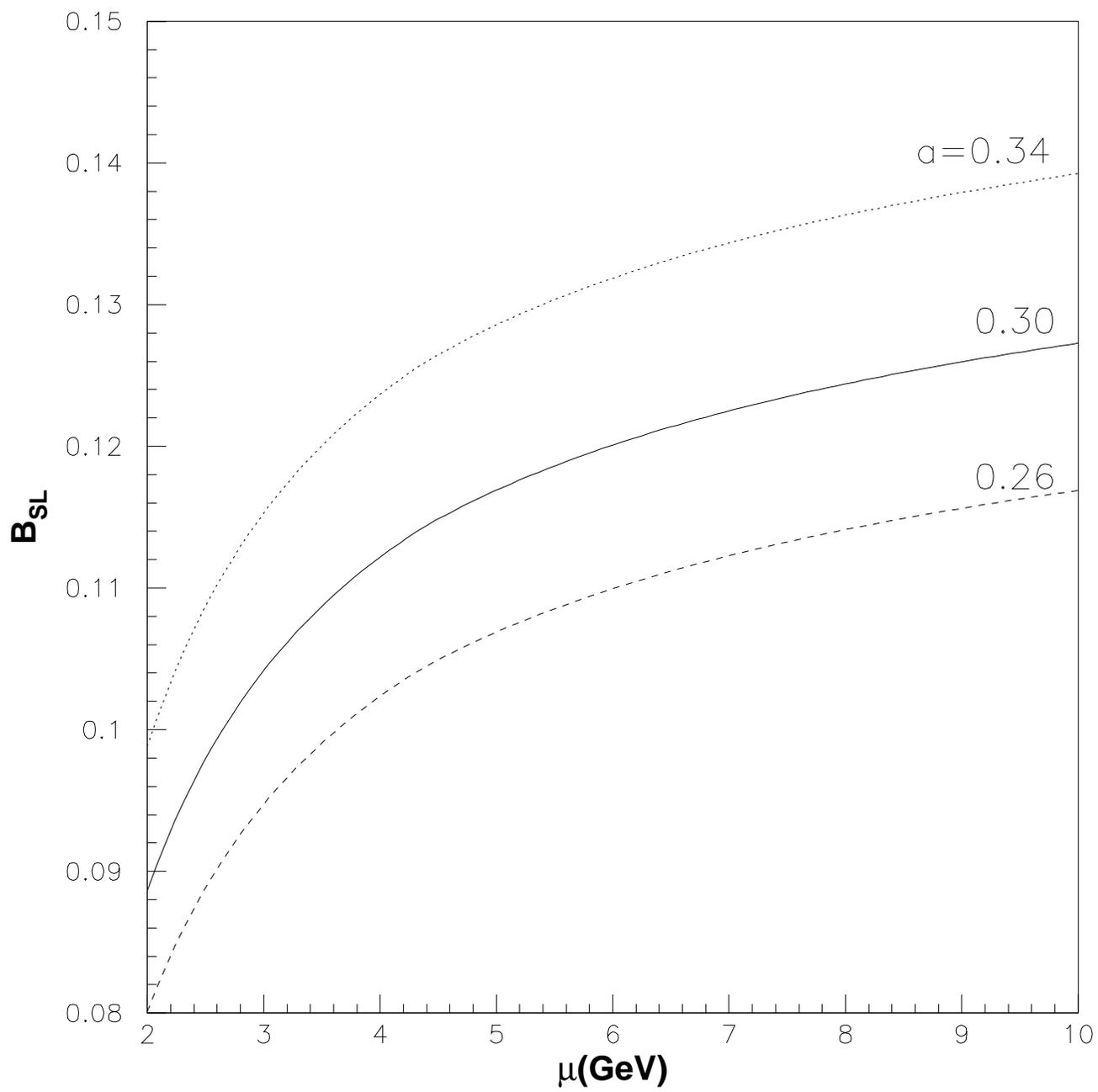
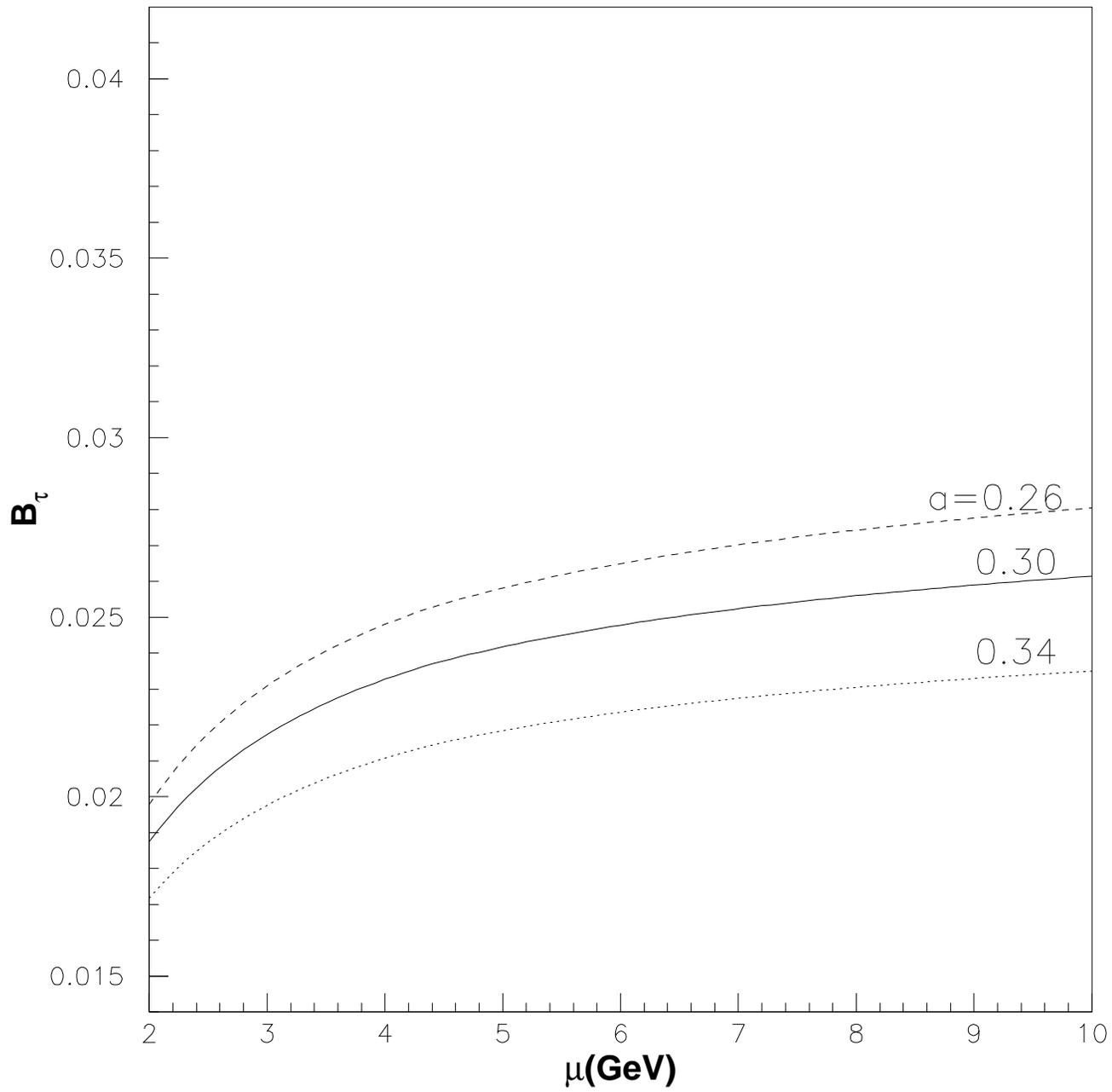
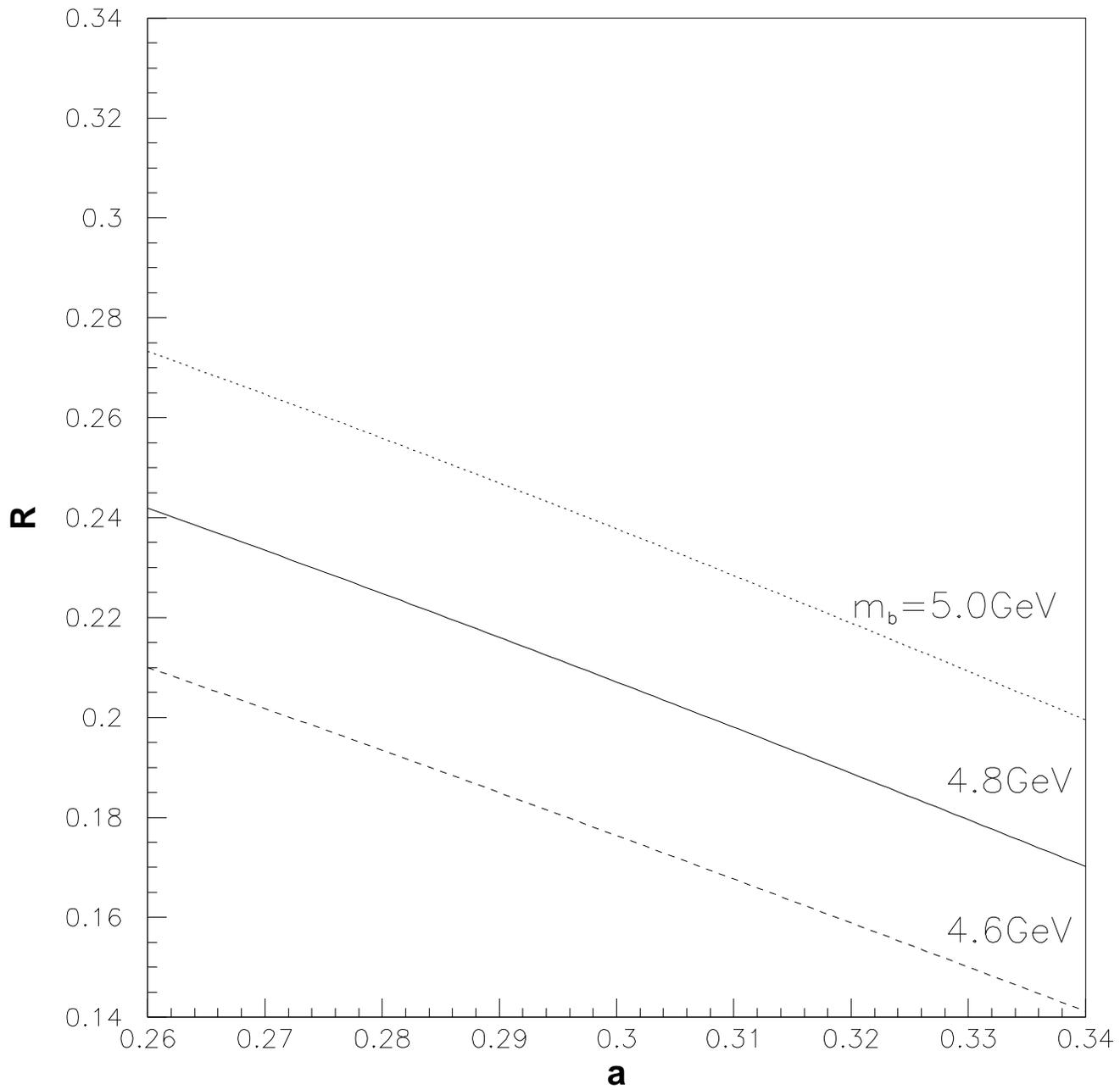
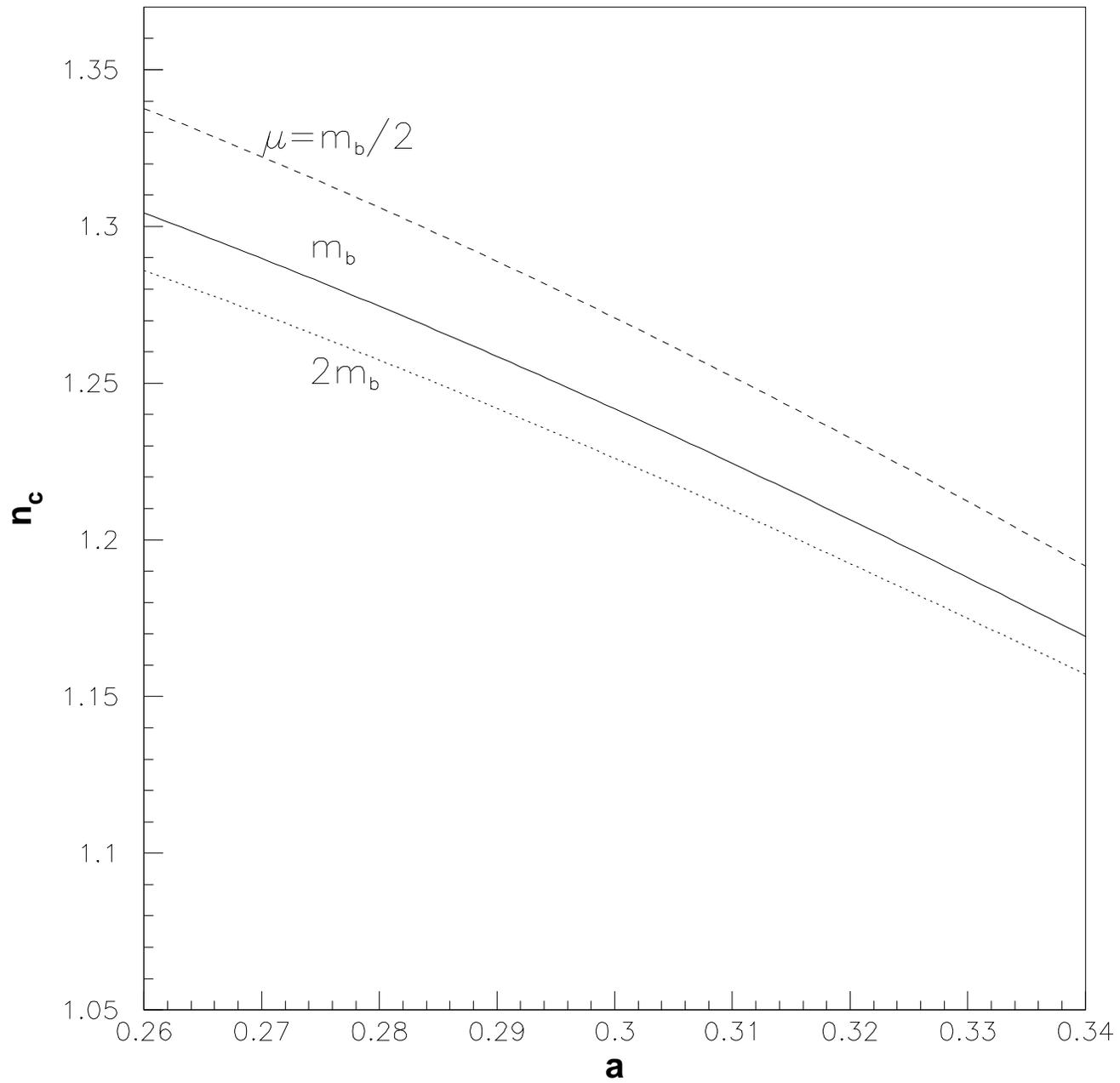
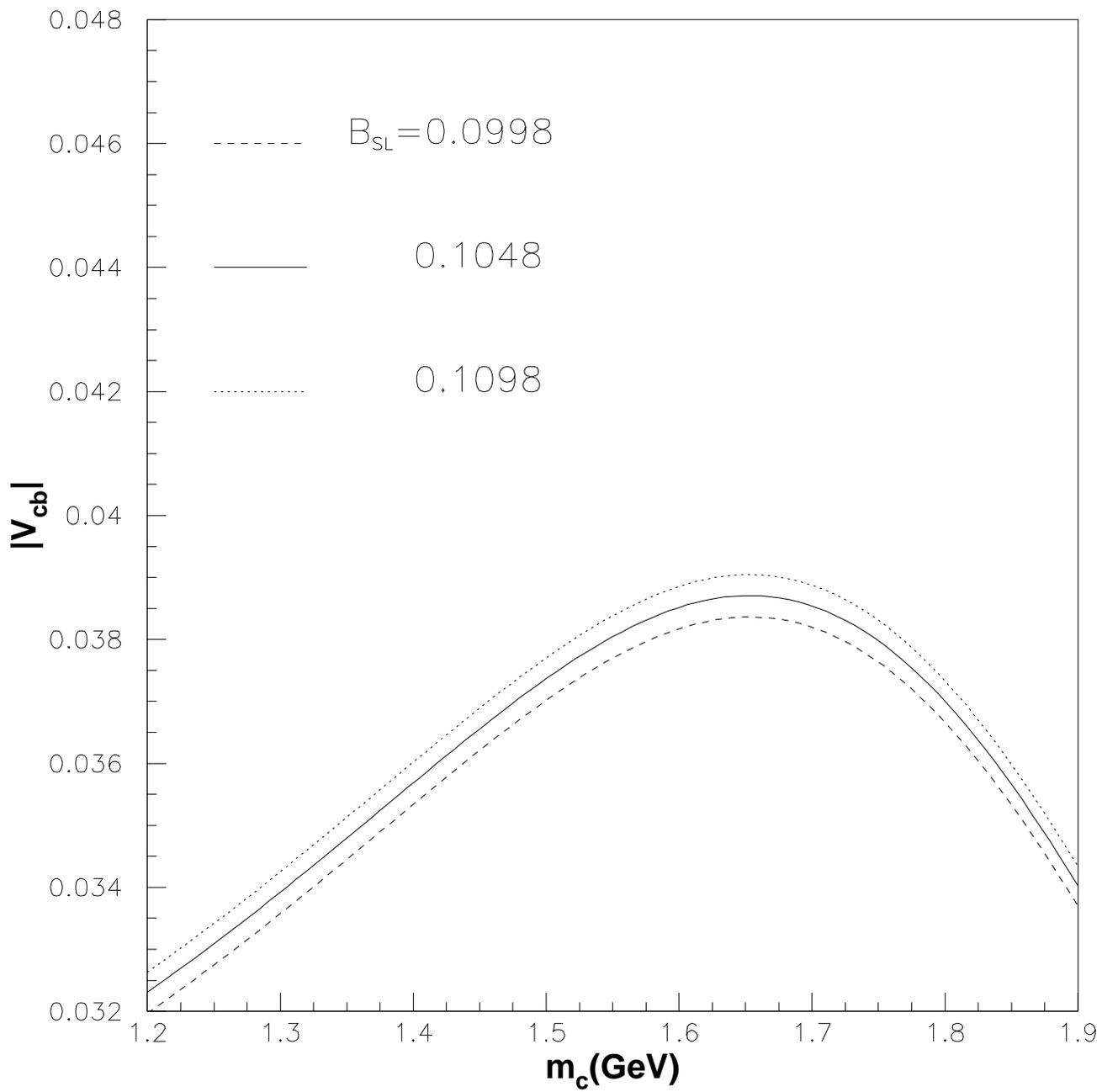
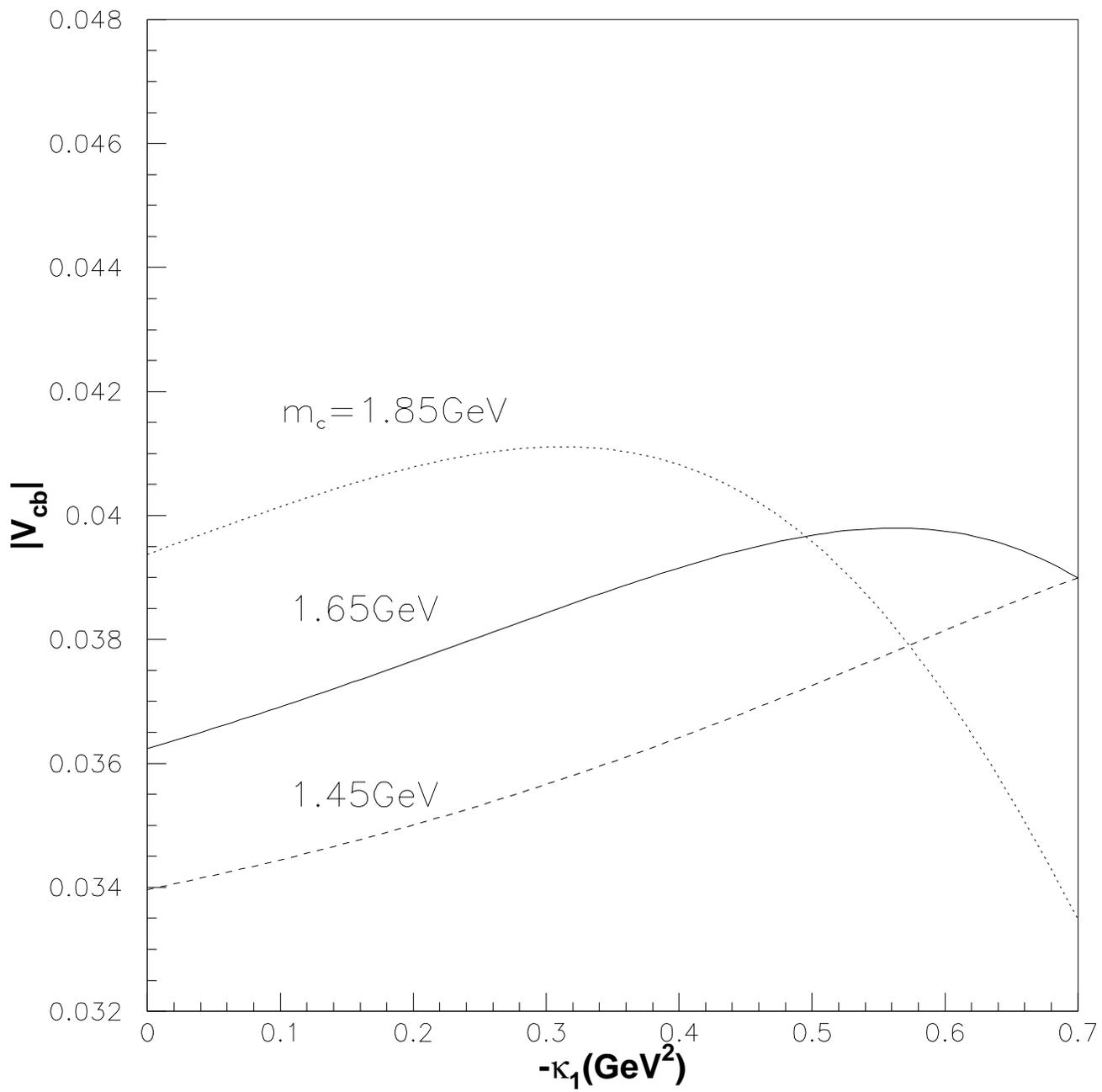
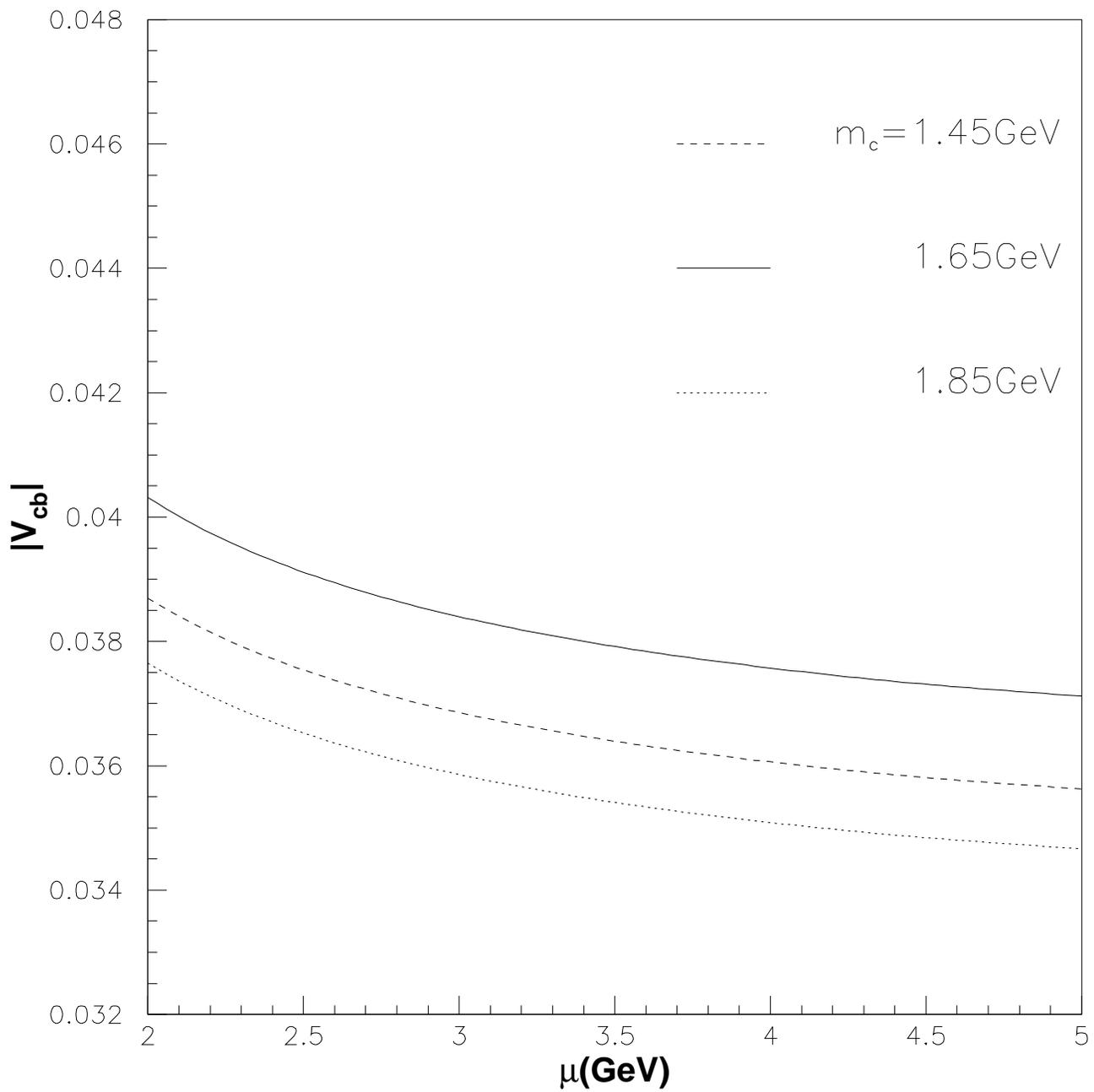
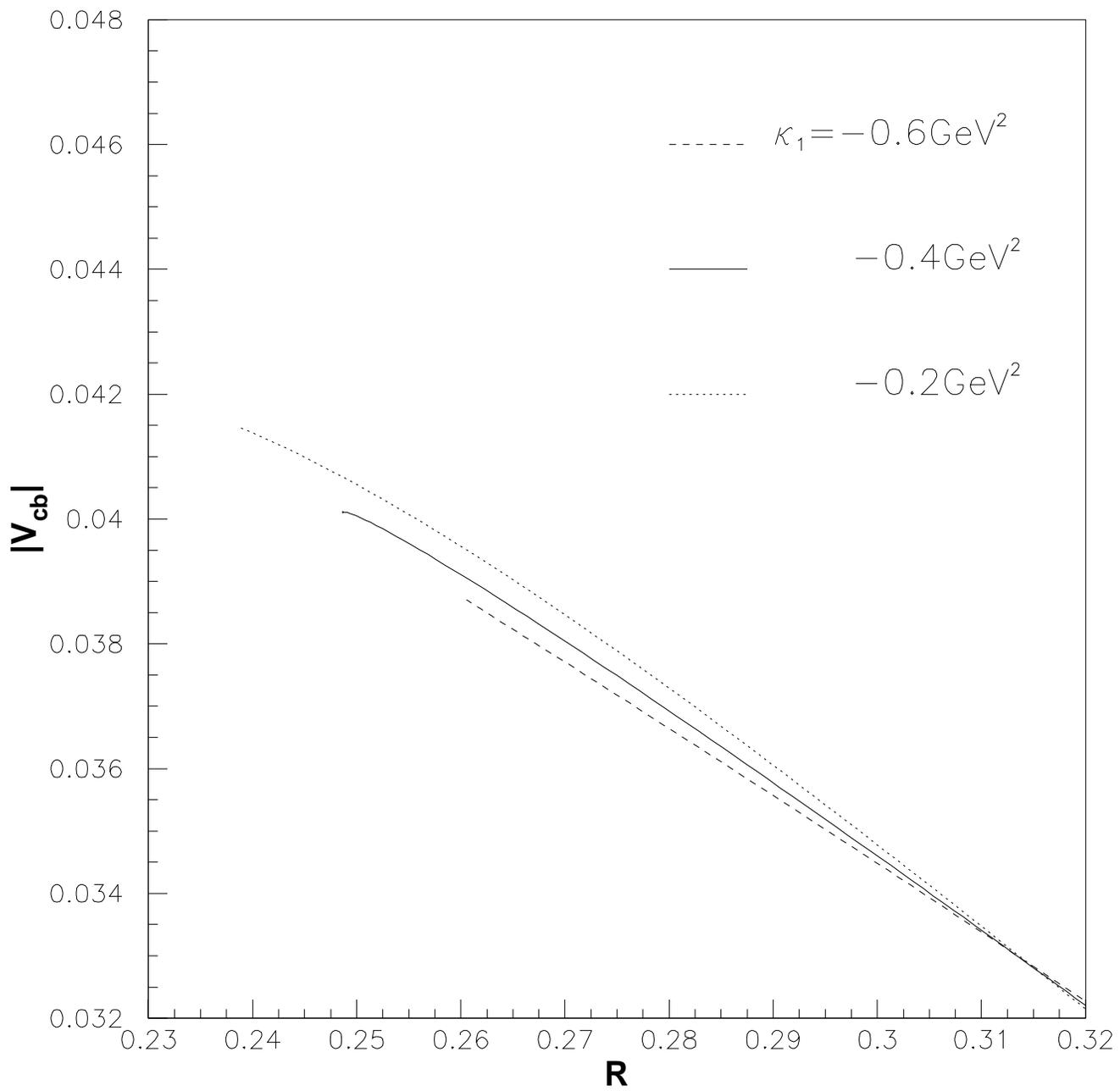
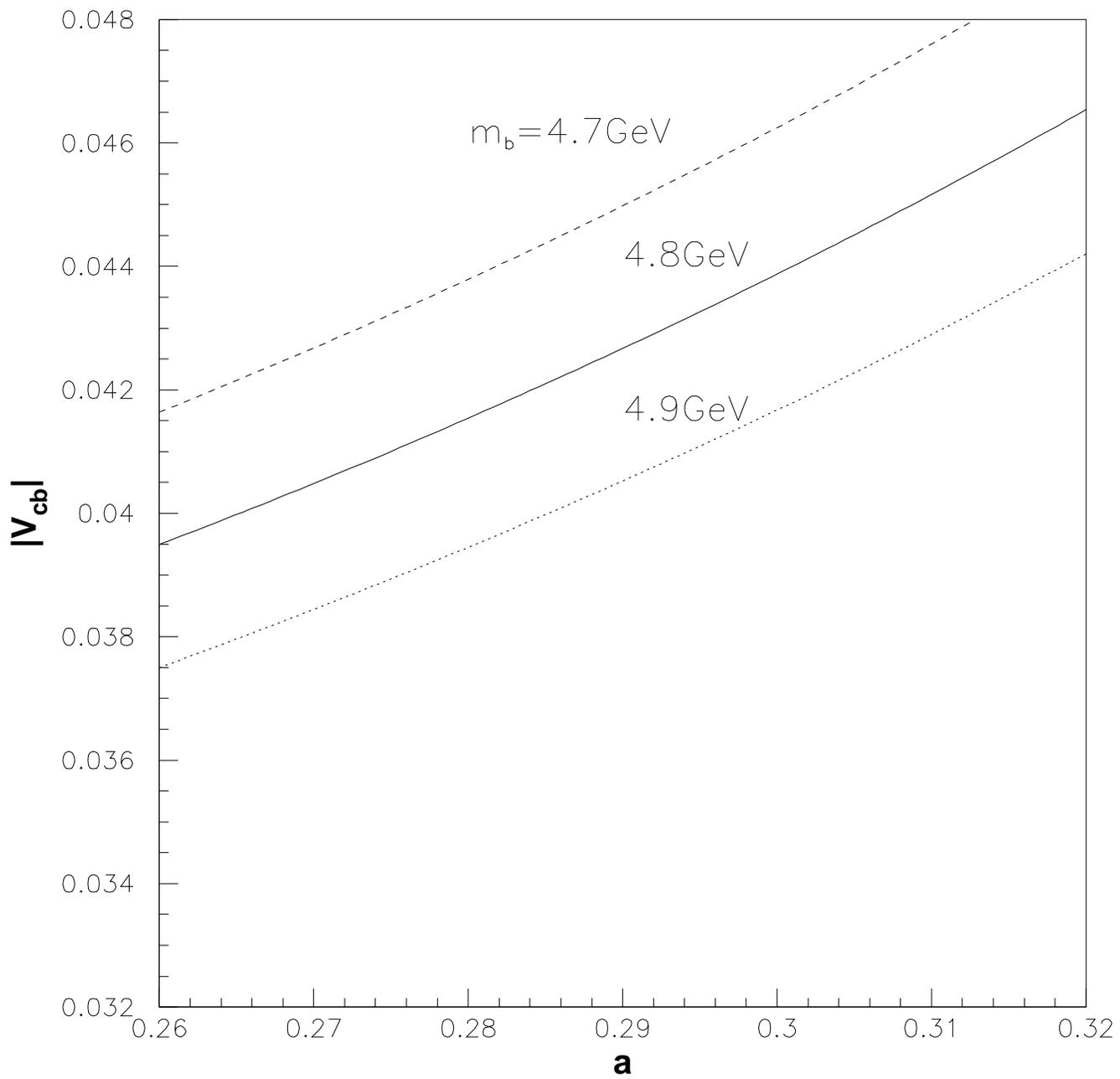
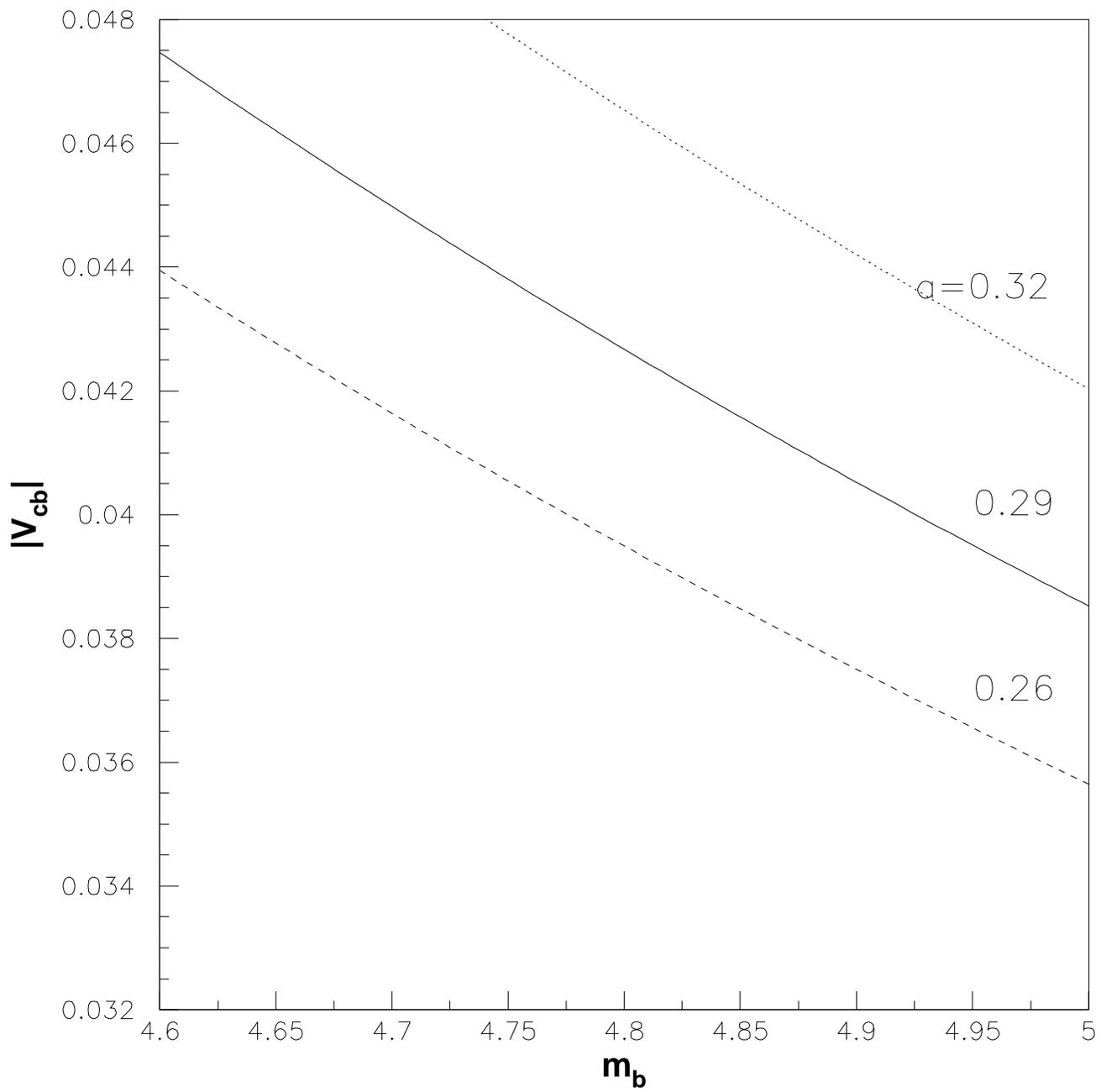
\end{document}